\documentclass[aps, preprint, nofootinbib, superscriptaddress, showpacs]{revtex4-1}

\usepackage{dcolumn}
\usepackage{bm}
\usepackage{color}
\usepackage{amsmath}
\usepackage{amssymb}
\usepackage{graphicx}% Include figure files
\usepackage{url}
\usepackage[colorlinks=false]{hyperref}
\usepackage[textsize=footnotesize]{todonotes}
\usepackage{lineno}
\usepackage[utf8]{inputenc}
\usepackage{graphicx}
\usepackage{geometry}
\usepackage{bm}
\usepackage{multirow}

\usepackage{subfig}

\def\eq#1{Eq.~(\ref{#1})}
\def\fig#1{Fig.~\ref{#1}}
\def\tab#1{Table.~\ref{#1}}
\def\eq#1{Eq.~(\ref{#1})}
\def\fig#1{Fig.~\ref{#1}}
\def\ket#1{|#1\rangle}

\def\IP#1#2{\langle#1|#2\rangle}
\def\BK#1#2#3{\langle#1|#2|#3\rangle}

\def\Xsgp{X^2\Sigma_g^+}
\def\Apiu{A^2\Pi_u}
\def\Bsup{B^2\Sigma_u^+}
\def\Csup{C^2\Sigma_u^+}
\def\Dpig{D^2\Pi_g}
\def\phg{^2\Phi_g}
\def\piu{^2\Pi_u}
\def\phu{^2\Phi_u}

\def\apiu{a^4\Pi_u}
\def\bsgm{b^4\Sigma_g^-}
\def\csup{c^4\Sigma_u^+}
\def\dsum{^4\Sigma_u^-}
\def\esum{^4\Sigma_u^+}
\def\fpig{f^4\Pi_g}
\def\hsgp{^4\Sigma_g^+}

\def\Init{\textrm{N}_2^+}% ionized nitrogen molecule

\def\muxy{\mu_\perp}
\def\muz{\mu_\parallel}
\def\muxyarrow{\stackrel{\muxy}{\longleftarrow}}
\def\muzarrow{\stackrel{\muz}{\longleftarrow}}

\begin{document}
\title{Disentangle pathways in strong field molecular photoionization by angular distribution of dissociation fragments}
\author{Xiangxu Mu$^*$}
\affiliation{State Key Laboratory for Mesoscopic Physics and Collaborative Innovation Center of Quantum Matter, School of Physics, Peking University, Beijing 10087, China}
\author{Ming Zhang}
\email{These authors contributed equally to this work.}
\affiliation{State Key Laboratory for Mesoscopic Physics and Collaborative Innovation Center of Quantum Matter, School of Physics, Peking University, Beijing 10087, China}
\author{Hanwei Yang}
\affiliation{State Key Laboratory for Mesoscopic Physics and Collaborative Innovation Center of Quantum Matter, School of Physics, Peking University, Beijing 10087, China}
\author{Haitan Xu}
\email{xuht@sustech.edu.cn}
\affiliation{Shenzhen Institute for Quantum Science and Engineering, Southern University of Science and Technology, Shenzhen 518055, China}
\affiliation{School of Physical Sciences, University of Science and Technology of China, Hefei 230026, China}
\author{Song Bin Zhang}
\email{song-bin.zhang@snnu.edu.cn}
\affiliation{Department of Physics, Shaanxi Normal University, Xi'an 710119, People's Republic of China}
\author{Lushuai Cao}
\affiliation{Wuhan National Laboratory for Optoelectronics and School of Physics, Huazhong University of Science and Technology, Wuhan 430074, China}
\author{Min Li}
\affiliation{Wuhan National Laboratory for Optoelectronics and School of Physics, Huazhong University of Science and Technology, Wuhan 430074, China}
\author{Zijian L\"u}
\affiliation{State Key Laboratory for Mesoscopic Physics and Collaborative Innovation Center of Quantum Matter, School of Physics, Peking University, Beijing 10087, China}
\author{Chengyin Wu}
\affiliation{State Key Laboratory for Mesoscopic Physics and Collaborative Innovation Center of Quantum Matter, School of Physics, Peking University, Beijing 10087, China}
\author{Zheng Li}
\email{zheng.li@pku.edu.cn}
\affiliation{State Key Laboratory for Mesoscopic Physics and Collaborative Innovation Center of Quantum Matter, School of Physics, Peking University, Beijing 10087, China}

\date{\today}

\begin{abstract}
In strong field ionization, the pump pulse not only photoionizes the molecule, but also drives efficient population exchanges between {its ionic ground and excited states.}
In this study, we investigated the population dynamics accompanying strong field molecular photoionization, using angular distribution of  dissociative fragments after ionization.
Our results reveal that the first and higher order processes of the post-ionization population redistribution mechanism (PPRM) in the ion core can be disentangled and classified by {its} angle-resolved kinetic energy release (KER) spectra.
We demonstrate that the imprints of PPRM in the KER spectra can be used to determine the branching ratio of the population exchange pathways of different orders, by exploiting the pump intensity dependent variation of the spectra.
\end{abstract}

\maketitle
\section{Introduction}
The photoionization and dissociation of molecules induced by intense femtosecond laser pulse are fundamental physical processes of light-matter interaction{s}.
The dissociation dynamics can be resolved in detail by delayed ultrashort probe laser pulses for {a} series of delays between the pump and probe laser pulses, where the pump pulse excites or ionizes the molecule and initiates the dissociation~\cite{corlin15:043415,abanador20:043410,rudenko18:013418}.
Although the ionization and dissociation of few-electron molecules, such as H$_2$, have been exhaustively studied~\cite{jpb1995,prl1883,prl515,prl692,prl4876,prl163004,prl093001,prl193001,science312246,prl223201,prl023002,prl143004,jpb2005}, the strong field ionization and subsequent dynamics of many-electron molecules do not have a coherent and comprehensive characterization of the underlying physical processes due to the complexity of the involved electronic and nuclear degrees of freedom.\\
Different from the often adopted sudden ionization approximation, after the {leaving of} ionized electrons {from} the molecular ionic core in a few hundreds of attoseconds~\cite{robin11:053003,pra033402,pra033423}, the electronic configurations between the ionic states can still be significantly modified within the duration of the ionization pulse, because the laser {couples} all the relevant {ionic} states allowed by symmetry.
And the ionization pulses {with pulse duration of about 10 fs} can provide sufficient time for such laser-driven population dynamics.
The resulted post-ionization population distribution mechanism (PPRM) has proven to be important in many scenarios and applications of strong field physics, e.g. it is crucial for the realization of population inversion in the N$_2^+$ air laser system driven by intense near infrared laser~\cite{prl143007}.
Quantitatively resolving the dynamics of PPRM is thus indispensable to devise new control mechanisms in these applications.\\
In this paper, we show that the pathways of electronic transitions in PPRM can be disentangled using the kinetic energy release (KER) spectra dissociative ionization as a function of angle between alignment axis and linear polarization of the ionization pump pulse.
It is demonstrated that the pathways of population redistribution can be classified by the orders of transitions {in the ionic states}, and their branching ratios can be quantified by measuring angle dependent KER's of dissociative fragments of laser aligned molecule at various pump intensities.
The proposed theory can serve as a new recipe to analyse population dynamics of molecules in the laser field by means of angle-resolved dissociative ionization, e.g. using cold-target recoil-ion-momentum spectroscopy (COLTRIMS).

\section{Theoretical methods}
\subsection{Model for the kinetic energy release spectra}

Kinetic energy release (KER) spectra is a central quantity in dissociative ionization experiments for the analysis of electronic dynamics and structural evolution of molecules~\cite{abanador_characterization_2020,ergler_spatiotemporal_2006,bocharova_time-resolved_2011,de_tracking_2010}. 
%%%%%%
Because the tunneling ionization rate exponentially depends on the strength of laser electric field, the ionization {dominantly} occurs near the peak of the pump laser pulses.
{Supposing a linearly polarized laser is used with polarization in space-fixed $Z$ axis, and $\theta$ is the angle between the molecule-fixed $z$ axis and space-fixed $Z$ axis.}
In our model, the dissociation of the {ionic} diatomic molecule {is separated} into two processes.
Firstly, ionization {of a neutral molecule} by an ultrashort intense pump laser {only happens at the time of its peak intensity and} generates molecular ions in different {excited} electronic states within the Franck Condon approximation,
{and the $\theta$-dependent angular distributions of ionic molecules are described by MO-ADK theory~\cite{pra033402}, which correspond to ionization rate $W_\nu(\theta)$ distribution along $\theta$ in electronic states $\nu$.}
{Secondly, after the generation of molecular ions at its peak, the remaining half falling pulse triggers the transition and dissociation between the electronic states of different aligned ionic molecules along $\theta$~\cite{prl143007}.}
{The dissociative state transferred from initial state $\nu$ dissociates and generates the kinetic energy spectra $D_\nu(\theta,E)$ for the ionic molecule along  $\theta$, where $E$ is the kinetic energy of the fragments.}
Thus, the total {KER spectra can be} constructed incoherently by the two processes above {as}, 
% 
%%%%%%%%
\begin{equation}\label{eq:ker_tot}
    I_\mathrm{KER}(\theta,E) = \sum_{\nu} W_\nu(\theta) D_\nu(\theta,E)\ .
\end{equation}
%%%%%%%%%%%%%%%%%%%%

\subsection{Alignment-angle dependent dissociation}
To simulate the population dynamics and dissociative ionization induced by the ionization laser pulse, let us consider a diatomic molecular cation, which is produced by ionization at the peak of the pulse. The electronic states $\ket{i}$ and $\ket{j}$ with potential energy curves (PEC) $V_i(r)$ and $V_j(r)$ are coupled by linearly polarized laser pulse, when allowed by symmetry selection rules.
Denote the distance between two atoms in diatomic molecule to be $r$,
the Hamiltonian $\hat{H}$ of the molecular ionic core can be written as 
%%%%%%%%%%%%%%%%%%%%%%%%%
\begin{equation}
    \hat{H} = \hat{T} + \hat{H}' - i\hat{W}\ ,
\end{equation}
%%%%%%%%%%%%%%%%%%%%%%%%%
where $\hat{T}$ is kinetic energy operator and $\hat{W}=\eta \Theta(r-r_c)(r-r_c)^2$ is the complex absorbing potential (CAP)~\cite{beck_multiconfiguration_2000},
$\Theta(r-r_c)$ is step function, where $r_c$ is the position of the flux plane and is chosen such that the potentials become constant for $r>r_c$. 
The potential and laser-molecule interaction Hamiltonian $\hat{H}'$ could be written as 
%%%%%%%%%%%%%%%%%%%%%%%%%
\begin{equation}
    \hat{H}'=\left(
    \begin{array}{ccc}
    V_i & \vec{\mu}_{ij}\cdot\vec{E} & \cdots\\
    \vec{\mu}_{ij}\cdot\vec{E} & V_j & \cdots \\
    \vdots & \vdots & \ddots
    \end{array}
    \right)\ , 
\end{equation}
%%%%%%%%%%%%%%%%%%%%%%%%%
where $i,j$ label electronic states, $\vec{\mu}_{ij}$ are the transition dipole moments, $\vec{E}$ is electric field of the laser. $\theta$ denotes the angle between the polarization axis of the ionization laser and the aligned molecular axis. 
We solve the time-dependent Schr\"odinger equation (TDSE) of the nuclear wavepacket, using the multi-configuration time-dependent Hartree (MCTDH) approach~\cite{beck_multiconfiguration_2000,mctdh:package},
%
%%%%%%%%%%%%%%%%%%%%%%%%%
\begin{equation}\label{TDSE}
    i \frac{\partial}{\partial t}\left(
    \begin{array}{c}
    \Psi_i      \\
    \Psi_j      \\
    \vdots
    \end{array}
    \right)=\hat{H}\left(
    \begin{array}{c}
    \Psi_i      \\
    \Psi_j      \\
    \vdots
    \end{array}
    \right)\,.
\end{equation}
%%%%%%%%%%%%%%%%%%%%%%%%%
%
The flux analysis is able to give the intensity of wavepacket components of specific kinetic energies, which go through the flux plane at $r=r_c$.
The population of dissociated wavepacket is given by the $S$-matrix elements,
%%%%%%%%%%%%%%%%%%%%%%%%%
\begin{equation}
    |S_{\nu}(E_{\nu})|^2 = \frac{d}{dt} \BK{\phi_\nu^+}{\Theta}{\phi_\nu^+}\ ,
\end{equation}
%%%%%%%%%%%%%%%%%%%%%%%%%
where $\phi_\nu^+$ denotes the asymptotic wavefunction of electronic state $\ket{\nu}$.
And the corresponding $S$-matrix elements could be calculated using CAP as~\cite{jackle_timedependent_1996,beck_multiconfiguration_2000}
%%%%%%%%%%%%%%%%%%%%%%%%%
\begin{equation}
\label{eq:Smat2}
   |S_{\nu}(E_{\nu})|^2= \frac{1}{|\Delta (E_{\nu}) |^2} \iint dt d\tau \BK{ \Psi (t + \tau) }{ W } { \Psi (t) } \exp (-i E \tau) \ , 
\end{equation}
%%%%%%%%%%%%%%%%%%%%%%%%%
where 
%%%%%%%%%%%%%%%%%%%%%%%%%
\begin{equation}
    \Delta (E_{\nu}) =  \IP{\Psi(t)}{\Psi(t=0)}\ ,
\end{equation}
%%%%%%%%%%%%%%%%%%%%%%%%%
and $E_{\nu}$ is the eigenenergy of continuum components of the outgoing wavepacket. 
The alignment angle- and state-resolved kinetic energy release (KER) spectra $D_\nu(\theta,E)$ of dissociative ionization fragments could be thus obtained by \eq{eq:Smat2} for each electronic state $\ket{\nu}$, where the asympototic energy of each state $E^{\mathrm{as}}_{\nu}$ is subtracted, i.e. $E=E_{\nu}-E^{\mathrm{as}}_{\nu}$. 

\subsection{Alignment-angle dependent ionization}
Alignment angle dependent tunneling ionization probabilities $W_\nu(\theta)$ is modelled with the MO-ADK theory~\cite{pra033402, pra033423}. Tunneling ionization rate by intense laser is
%%%%%%%%%%%%%%%%%%%%%%
\begin{equation}
W_\nu(E, \theta)=\left(\frac{3 E}{\pi \kappa^{3}}\right)^{1 / 2} W_\mathrm{s t a t}(E, \theta)\ ,
\end{equation}
%%%%%%%%%%%%%%%%%%%%%%
where $E$ is the electric field strength, $\kappa=\sqrt{2I_p}$ and $I_p$ is the ionization potential of given molecular orbital, $\theta$ is the angle between molecular and laser polarization axis, and $W_\mathrm{stat}(E, \theta)$ is static tunneling ionization rate,
%%%%%%%%%%%%%%%%%%%%%%
\begin{eqnarray}
W_{\mathrm{stat }}(E, \theta)&=&\sum_{m^{\prime}}\frac{B^{2}\left(m^{\prime}\right)}{2^{\left|m^{\prime}\right|}\left|m^{\prime}\right| !}\frac{1}{\kappa^{2 Z_{c} / \kappa-1}} \\
& \times&\left(\frac{2 \kappa^{3}}{E}\right)^{2 Z_{c} / \kappa-\left|m^{\prime}\right|-1} e^{-2 \kappa^{3} / 3 E}\ ,
\end{eqnarray}
%%%%%%%%%%%%%%%%%%%%%%%
with $Z_{c}$ being the effective dissociative charge, and
%%%%%%%%%%%%%%%%%%%%%%%
\begin{equation}
B\left(m^{\prime}\right)=\sum_{l} C_{l} D_{m^{\prime}, m}^{l}(\theta) Q\left(l, m^{\prime}\right)\ ,
\end{equation}
%%%%%%%%%%%%%%%%%%%%%%%
where 
%%%%%%%%%%%%%%%%%%%%%%%
\begin{equation}
Q(l, m)=(-1)^{m} \sqrt{(\frac{(2 l+1)(l+|m|) )!}{2(l-|m|) !}}\ ,
\end{equation}
%%%%%%%%%%%%%%%%%%%%%%%
$D_{m^{\prime}, m}^{l}(\theta)$ is rotation matrix, and $C_l$ are the coefficients of molecular orbital given in~\cite{pra033402,pra033423}.
The laser pulse envelope is
%%%%%%%%%%%%%%%%%%%%%%%
\begin{equation}
    I(t) = I_0 \exp \left[ -\frac{1}{2}\left(\frac{t}{\tau}\right)^2\right]\ ,
\end{equation}
%%%%%%%%%%%%%%%%%%%%%%%
where $\tau$ is the pulse length at full width at half maximum (FWHM), which is set to be 40 fs in the simulation.
The ionization rate of an aligned molecule in state $\ket{\nu}$ is
%%%%%%%%%%%%%%%%%%%%%%%
\begin{equation}
    W_\nu(\theta) = \int_{-\infty}^{\infty} W_\nu(E(t),\theta) dt \ .
\end{equation}
%%%%%%%%%%%%%%%%%%%%%%%

\section{Results and Discussion}
In this section, we present the simulation of dissociative ionization of laser aligned N$_2$ and O$_2$ molecules, and the analysis of the post-ionization population dynamics of the molecular ion induced via PPRM of the ionization pulse.
%%%%%%
For all numerical applications discussed in this paper, we consider laser pulse with a fixed wavelength of 800 nm with a full width at the amplitude half maximum (FWHM) of 40 fs.
The center of the pulse is chosen to be at $t=0$ fs, which corresponds to the instant ionization at the peak of the laser at $t=0$ fs. The population exchange and dissociation dynamics takes place in the falling edge of the Gaussian pulse.

\begin{figure}[htb!]
    \centering
    \includegraphics[width=0.7\textwidth]{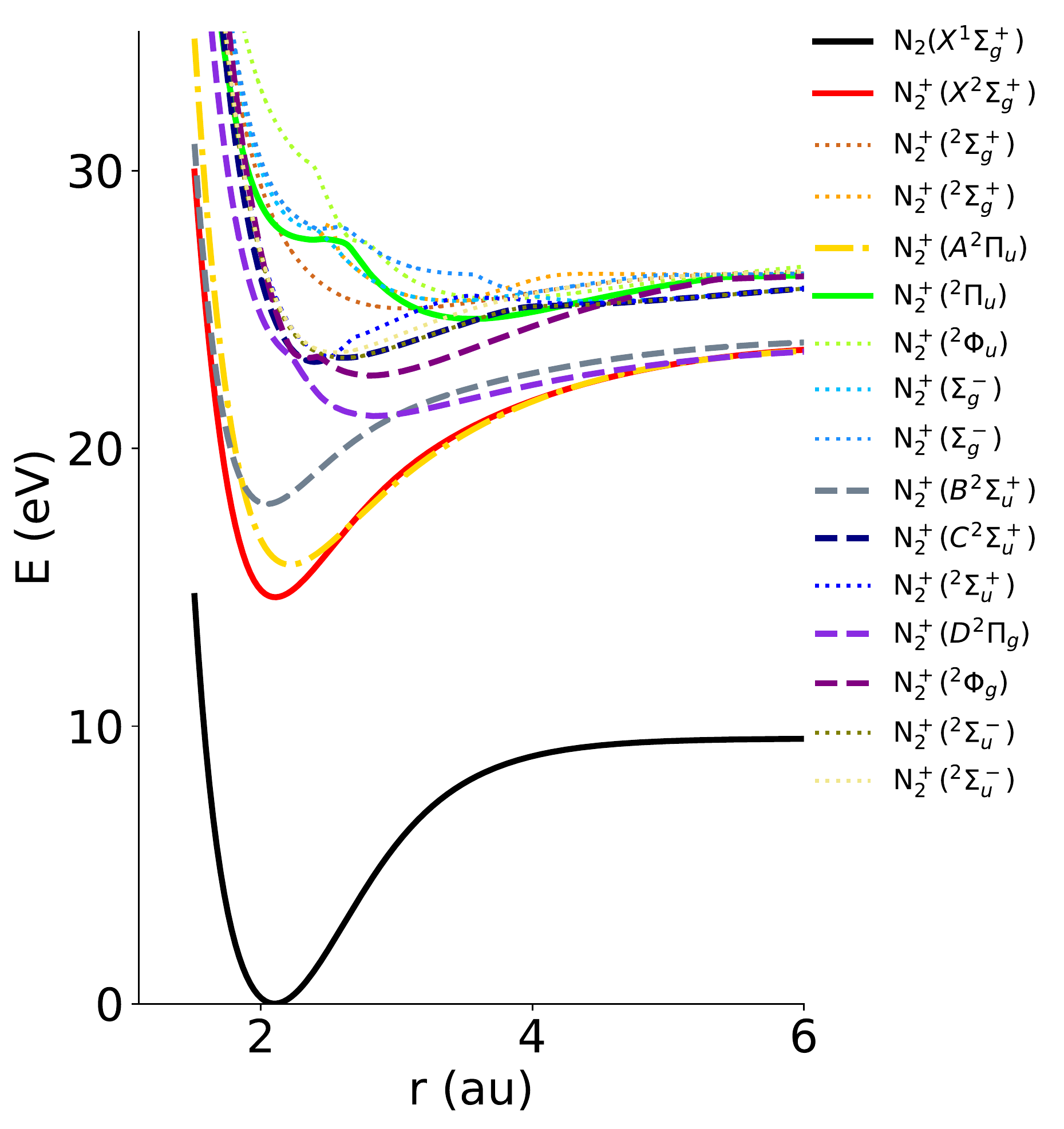}
    \caption{The potential energy curves of electronic states of N$_2^+$. Red line is for the ground cationic state $\Xsgp$, yellow line is for the dissociative state $\piu$.}
    \label{fig:pec}
\end{figure}

Via ab initio quantum chemistry calculations by the MOLPRO package, we constructed accurate PECs of nitrogen molecule that consist of the three low-lying electronic states directly accessible by ionization ($\Xsgp$, $\Apiu$ and $\Bsup$), the dissociated state $\piu$, and intermediate states $\Bsup$, $\Csup$, $\Dpig$ and $\phg$ of $\Init$, and the ground state $X^1\Sigma_g^+$ of N$_2$.
For the oxygen molecule, we calculated the PECs for the low-lying electronic state $\apiu$, dissociated state $\fpig$ and intermediate states $\bsgm$, $\csup$, $\dsum$, $\esum$, $\fpig$ and $\hsgp$ of O$_2^+$, and the ground state $^3\Sigma_g^+$ of O$_2$. 
Specifically, Dunning's aug-cc-pVTZ basis, active space including 14 valence orbitals of N atoms and O atoms were employed in the complete active space self-consistent-field (CASSCF) calculations. 
The calculated PECs are presented in \fig{fig:pec} and \fig{fig:o2pec}.
The initial population distribution of $\Init$ in the lowest three lowest electronic states are set to be 0.6 ($\Xsgp$), 0.2 ($\Apiu$) and 0.2 ($\Bsup$)~\cite{prl143007}.
Because of fast decoherence induced by the outgoing photoelectron within 1 fs ~\cite{robin11:053003}, the coherence between the $\Xsgp$, $\Apiu$ and $\Bsup$ states are neglected in our model.
For nitrogen cation, $\Xsgp$, $\Apiu$, $\Bsup$, $\Dpig$, $\phg$ are bounded states and $\piu$ is a dissociative state, which correlates to another state $\phu$ at $r=2.64$ a.u, although the $\Csup$ state is predissociative~\cite{ste20:032802}, its contribution to the KER spectra is negligibly small comparing to the direct dissociation channel in $\phu$ state, as confirmed by the MCTDH simulation.
The time-dependent wavefunctions in \eq{TDSE} are obtained by using the multi-configuration time-dependent Hatree (MCTDH) method~\cite{mctdh:package,beck_multiconfiguration_2000}. A complex absorbing potential (CAP) of the form $-iW(q)=-i\eta(r-r_0)^3$ was applied to calculate the photodissociation through flux analysis ~\cite{jackle_timedependent_1996}, with intensity $\eta=0.03$ a.u. and $r_0 = 10$ a.u. 512 sine-DVR basis functions are used for the $r$ grid.
In the simulation we include all the possible alignement angles of the molecular axis relative to the linear polarization of the ionization laser.
In the following sections, we investigate angle-resolved ionization rate of nitrogen molecule and the angle-resolved KER spectra $D(\theta,E)$ of dissociated fragments, as well as their dependence on peak intensity of the ionization pulse, they together provide the route towards disentangling the different pathways of population exchange.

%%%%%%

\subsection{Ionization of N$_2$ molecule}

\begin{figure}[htb!]
  \subfloat[]{\includegraphics[width=0.33\textwidth]{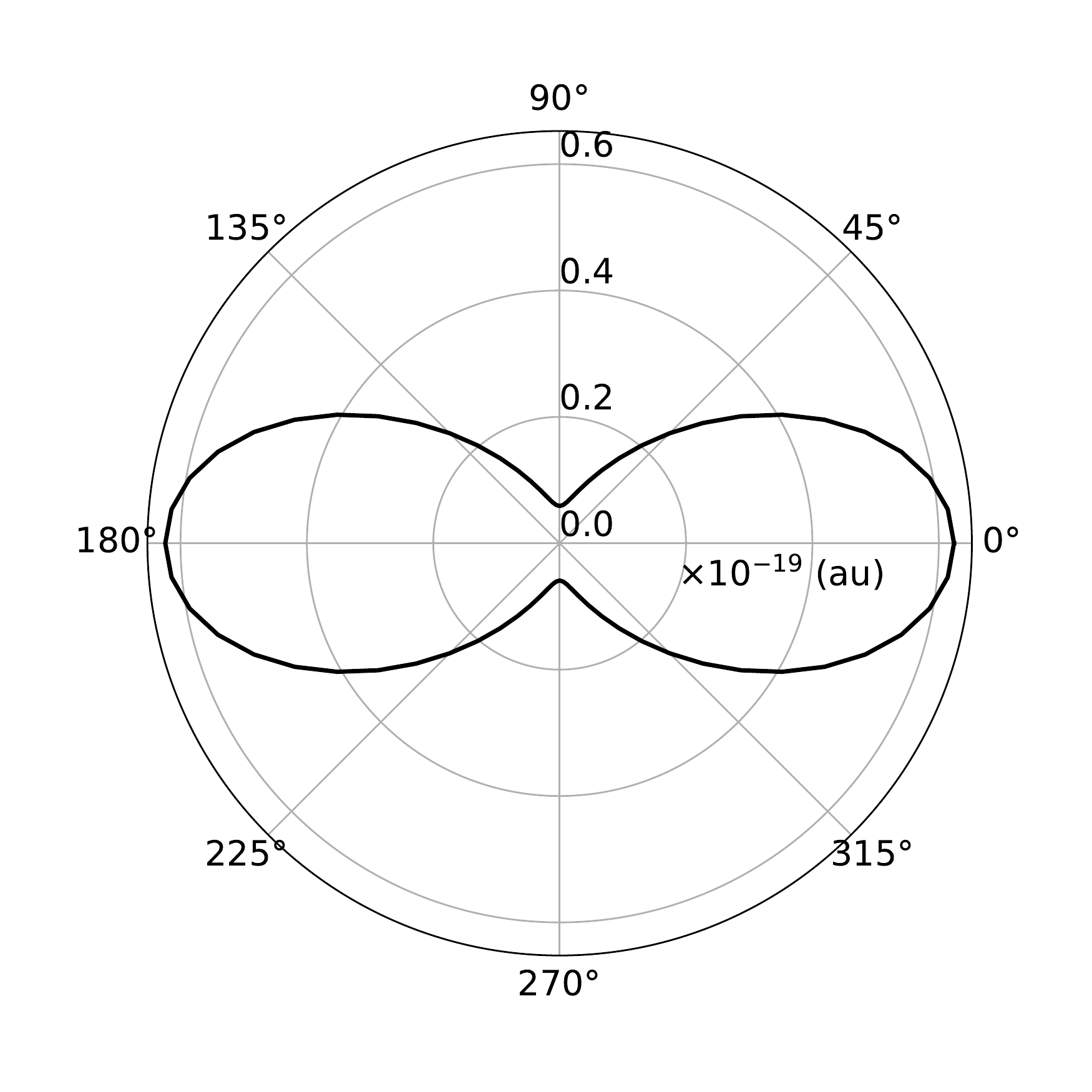}}
 \hfill 	
  \subfloat[]{\includegraphics[width=0.33\textwidth]{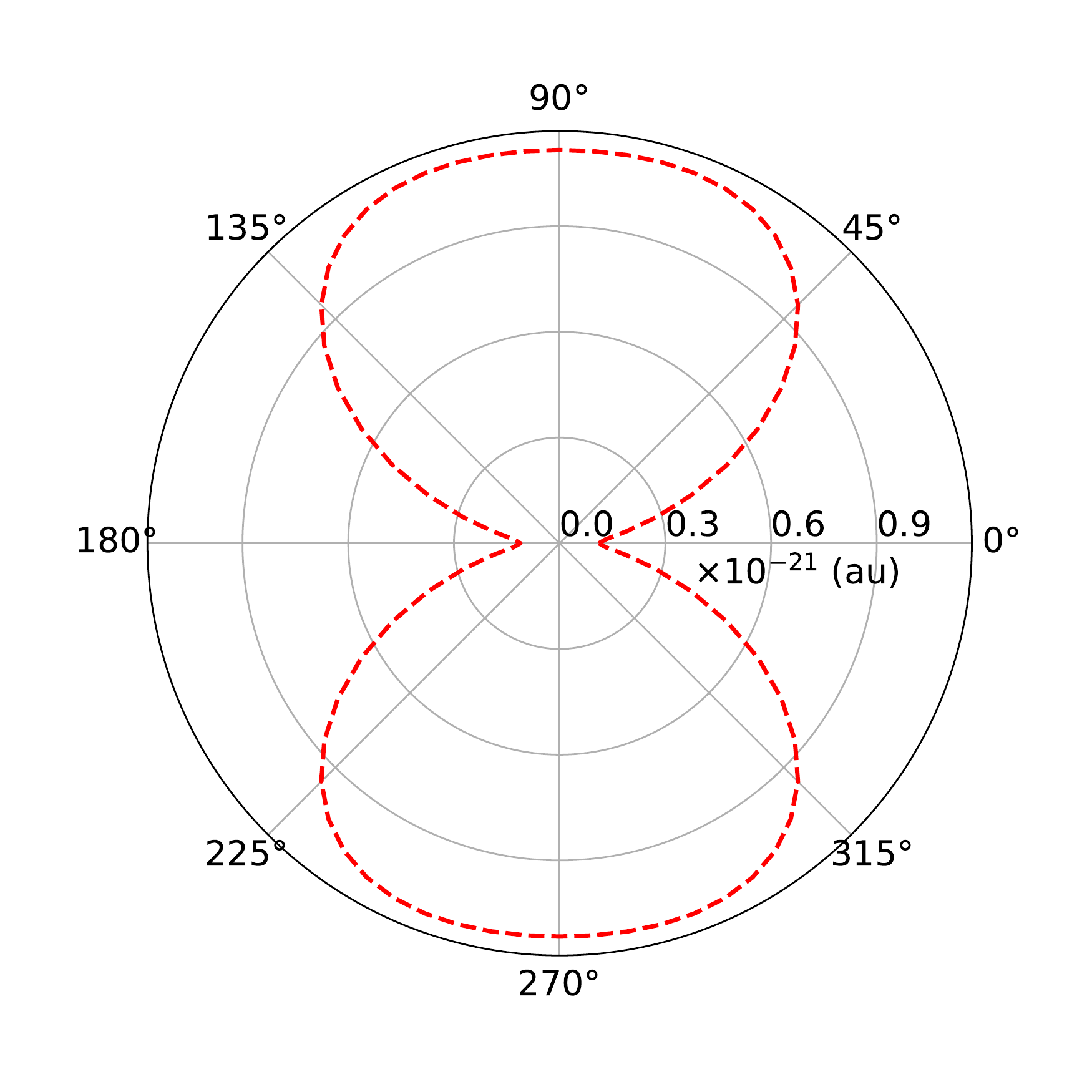}}
 \hfill	
  \subfloat[]{\includegraphics[width=0.33\textwidth]{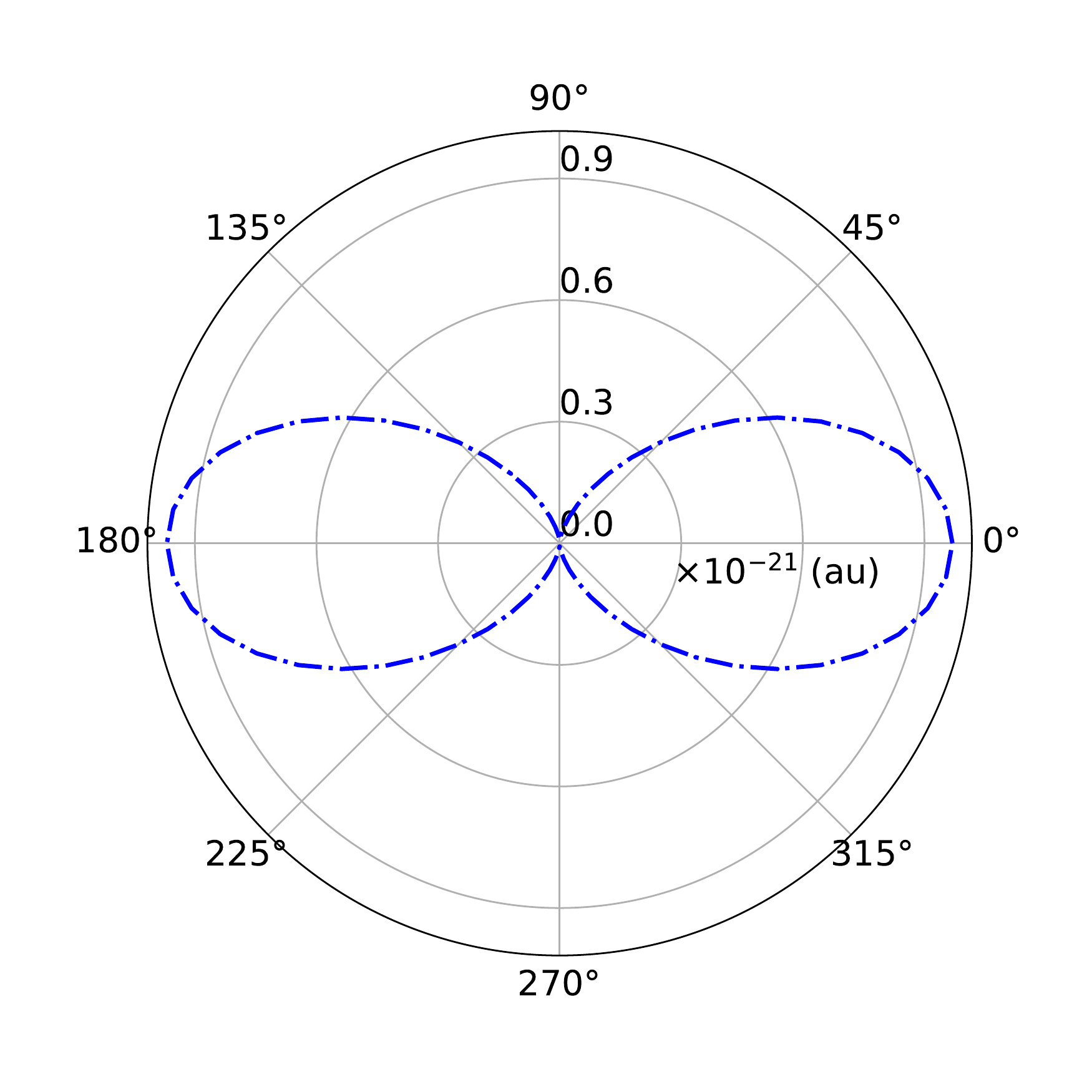}}
  \newline
  \subfloat[]{\includegraphics[width=0.33\textwidth]{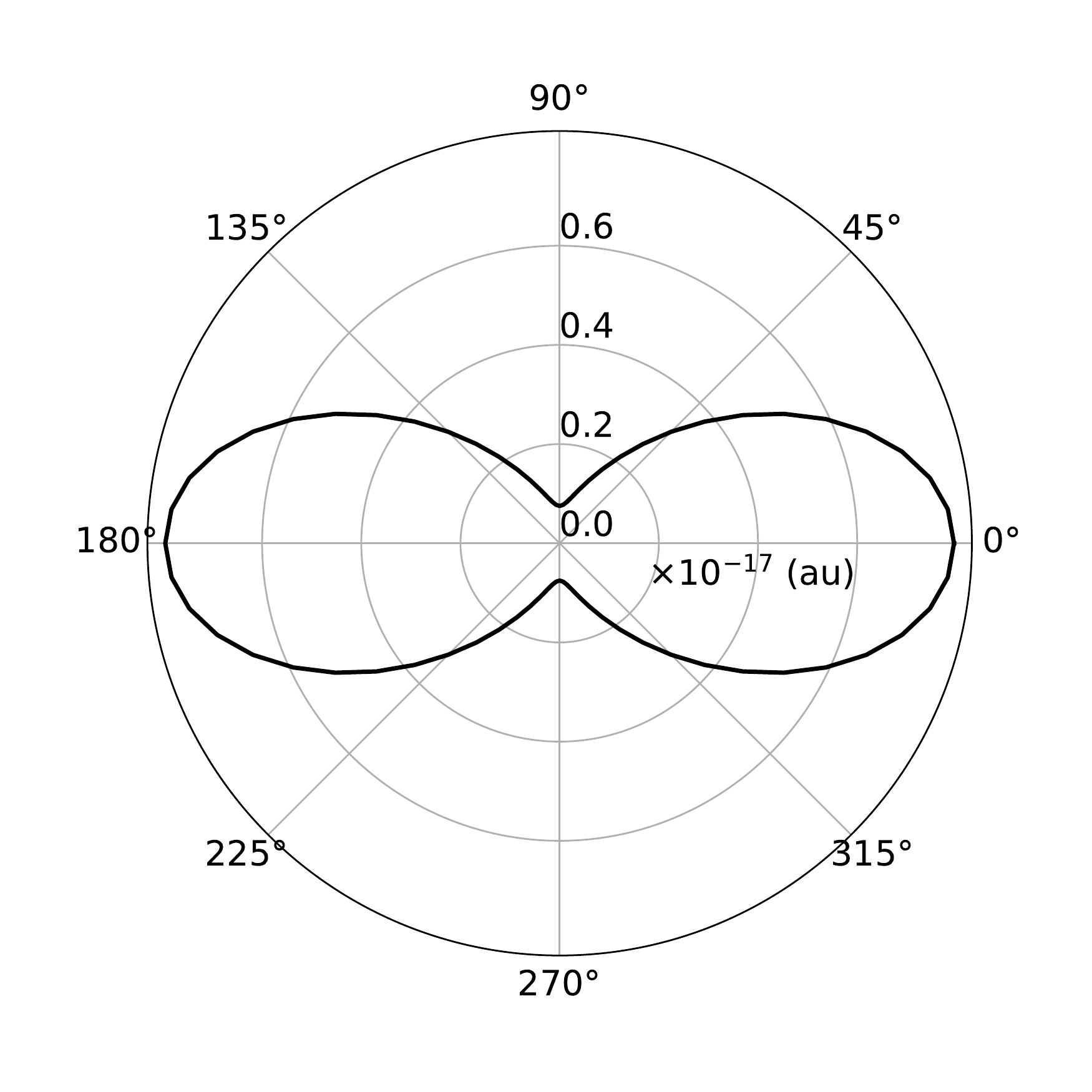}}
 \hfill 	
  \subfloat[]{\includegraphics[width=0.33\textwidth]{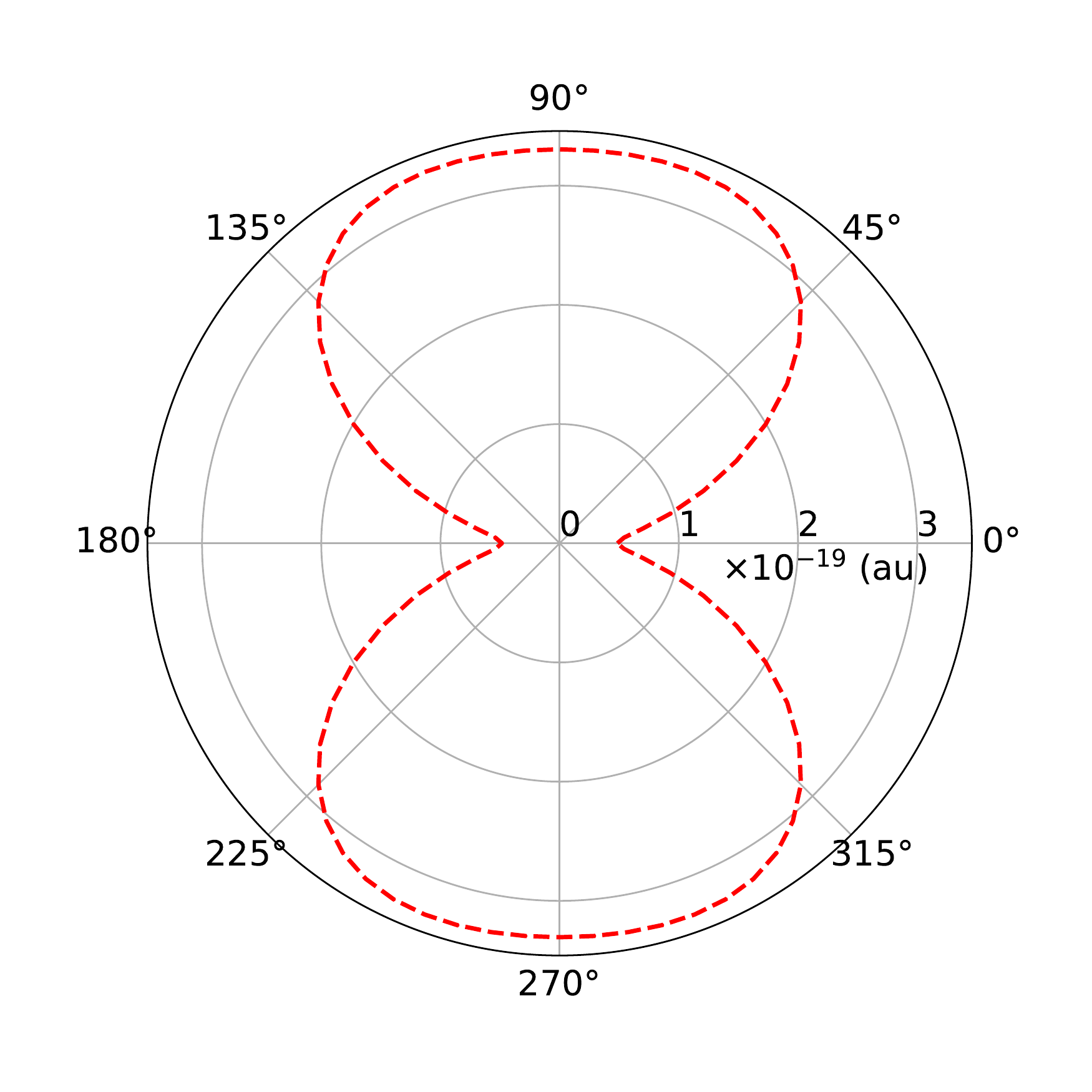}}
 \hfill	
  \subfloat[]{\includegraphics[width=0.3\textwidth]{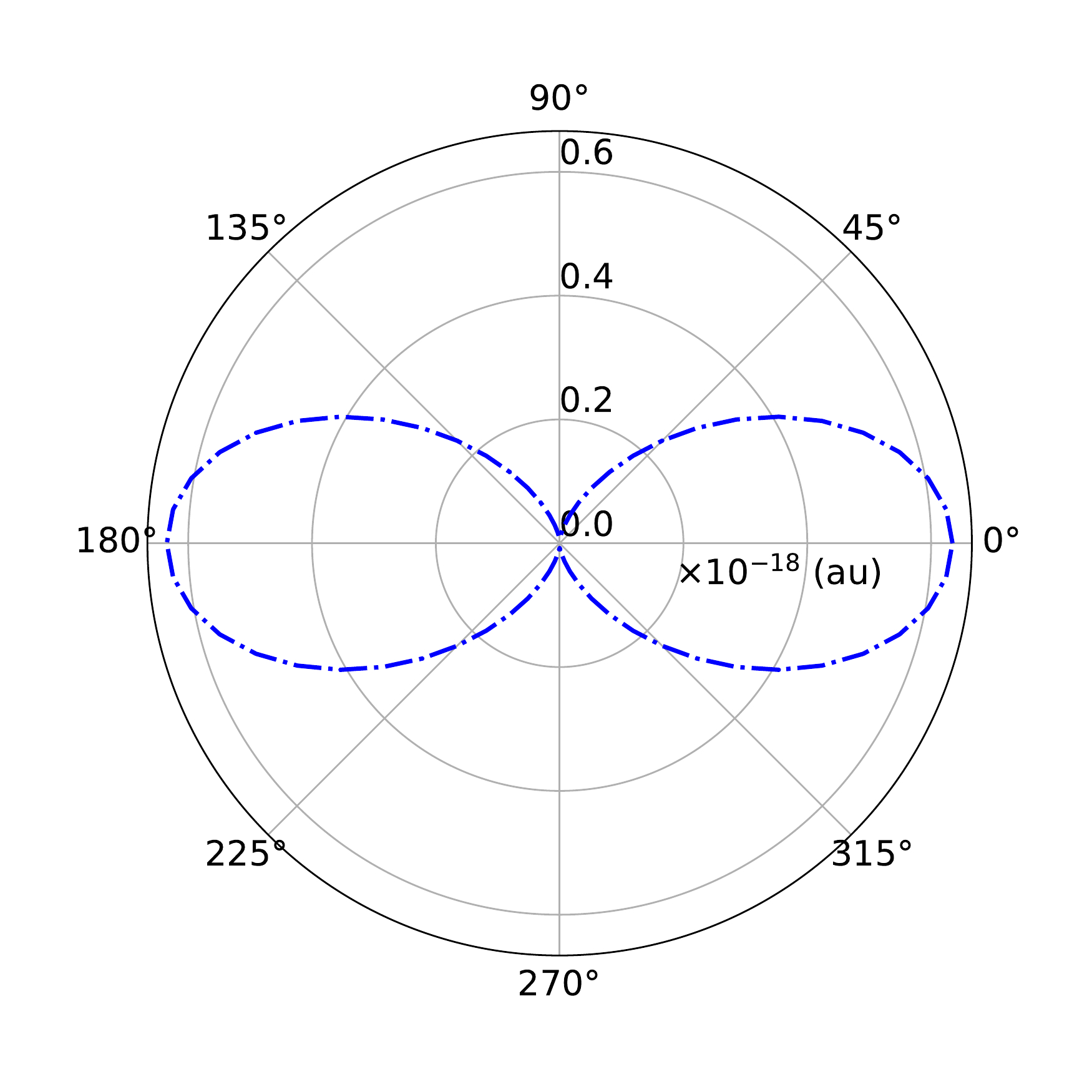}}
\caption{Tunneling ionization rate of $\Init$ at laser intensity of $1.0\times10^{14}$ W/cm$^2$ (first row) and $2.2\times10^{14}$ W/cm$^2$ (second row). Black solid line, red dashed line and blue point dashed line denotes the ionization rate of state $\Xsgp$, state $\Apiu$ and state $\Bsup$.}
\label{fig:moadk}
\end{figure}

In this section we calculate ionization rate for two representative ionization laser peak intensities $I_0$ (1.0 and 2.2$\times10^{14}$ W/cm$^2$) and the alignment angle dependent ionization rate is shown in \fig{fig:moadk}. 
Ionization rate of state $\Xsgp$ and $\Bsup$ maximize when molecule alignment direction is parallel to laser and minimize when molecule alignment direction is perpendicular to laser.
But ionization rate of state $\Apiu$ maximize at perpendicular alignment and minimize at parallel alignment. 
These results is consistent with the conclusion in Ref.~\cite{pra033423}.
%

%%%%%%%%%%%%%%%%%%%%%%%%%%%%%%%%%%%%%%
%%%%%%%%%%%%%%%%%%%%%%%%%%%%%%%%%%%%%%
%%%%%%%%%%%%%%%%%%%%%%%%%%%%%%%%%%%%%%
\begin{figure}[htb!]
    \centering
    \includegraphics[width=0.6\textwidth]{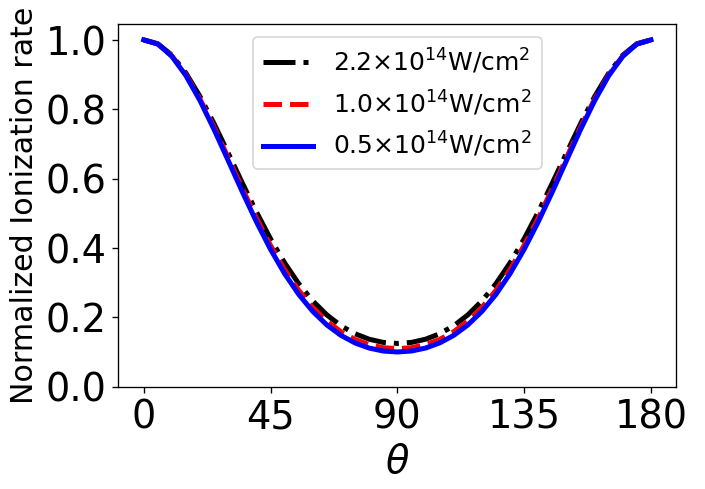}
    \caption{Normalized distribution of total ionization rate to final states $\Xsgp$, $\Apiu$ and $\Bsup$ as a function of angle between alignment axis and laser polarization. Black dotted dashed line, red dashed line and blud solid line denote the distribution under laser intensity of $2.2\times10^{14}$ W/cm$^2$, $1.0\times10^{14}$ W/cm$^2$ and $0.5\times10^{14}$ W/cm$^2$, respectively. }
    \label{fig:moadk2}
\end{figure}
%%%%%%%%%%%%%%%%%%%%%%%%%%%%%%%%%%%%%%

\subsection{Dissociation dynamics of N$_2^+$ cation}
%%%%%%%%%%%%%%%%%%%%%%%%%%%%%%%%%%
\begin{figure}[htb!] 	 	
  \subfloat[]{\includegraphics[height=4.12cm]{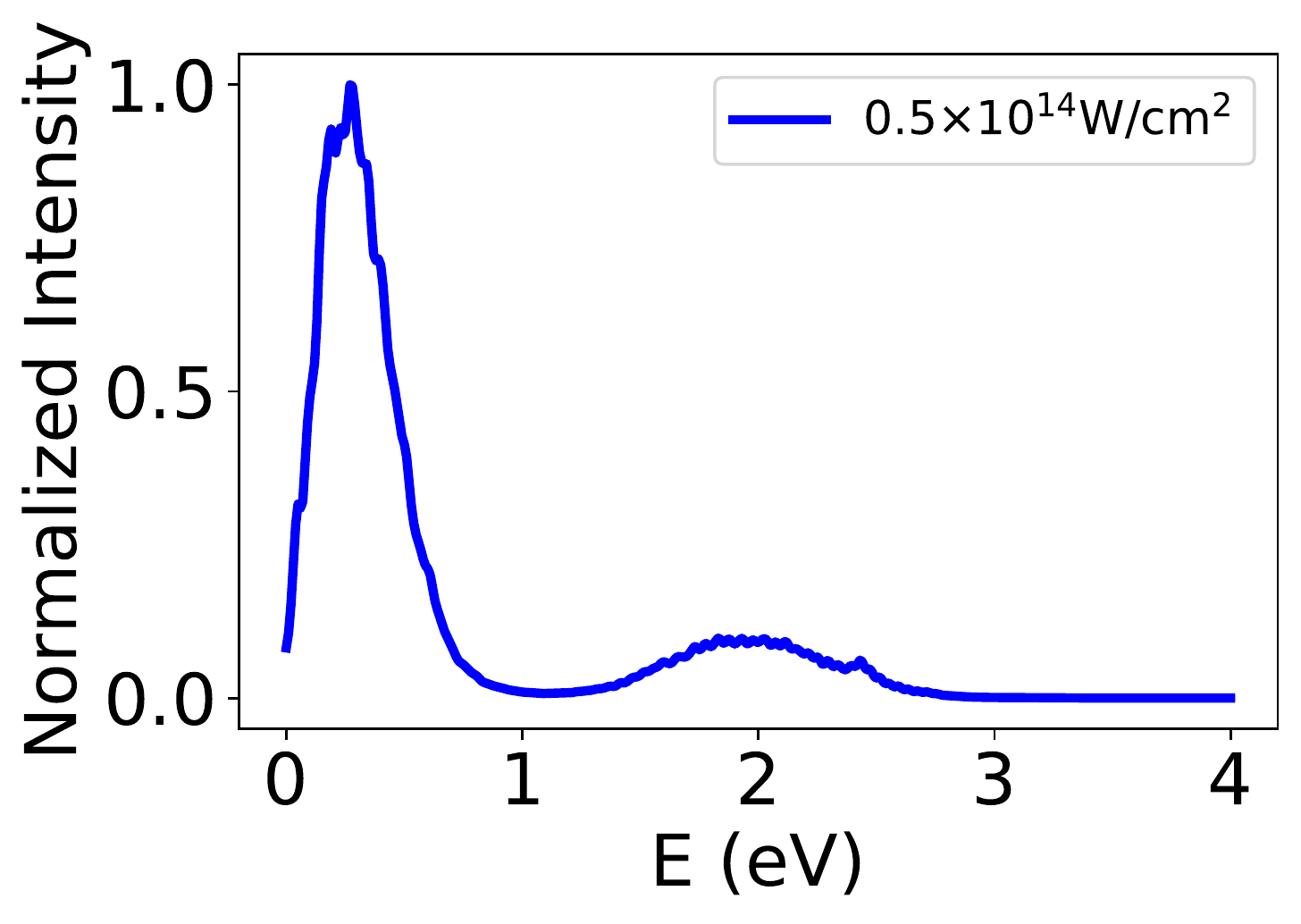}}
 \hfill 	
  \subfloat[]{\includegraphics[height=4cm]{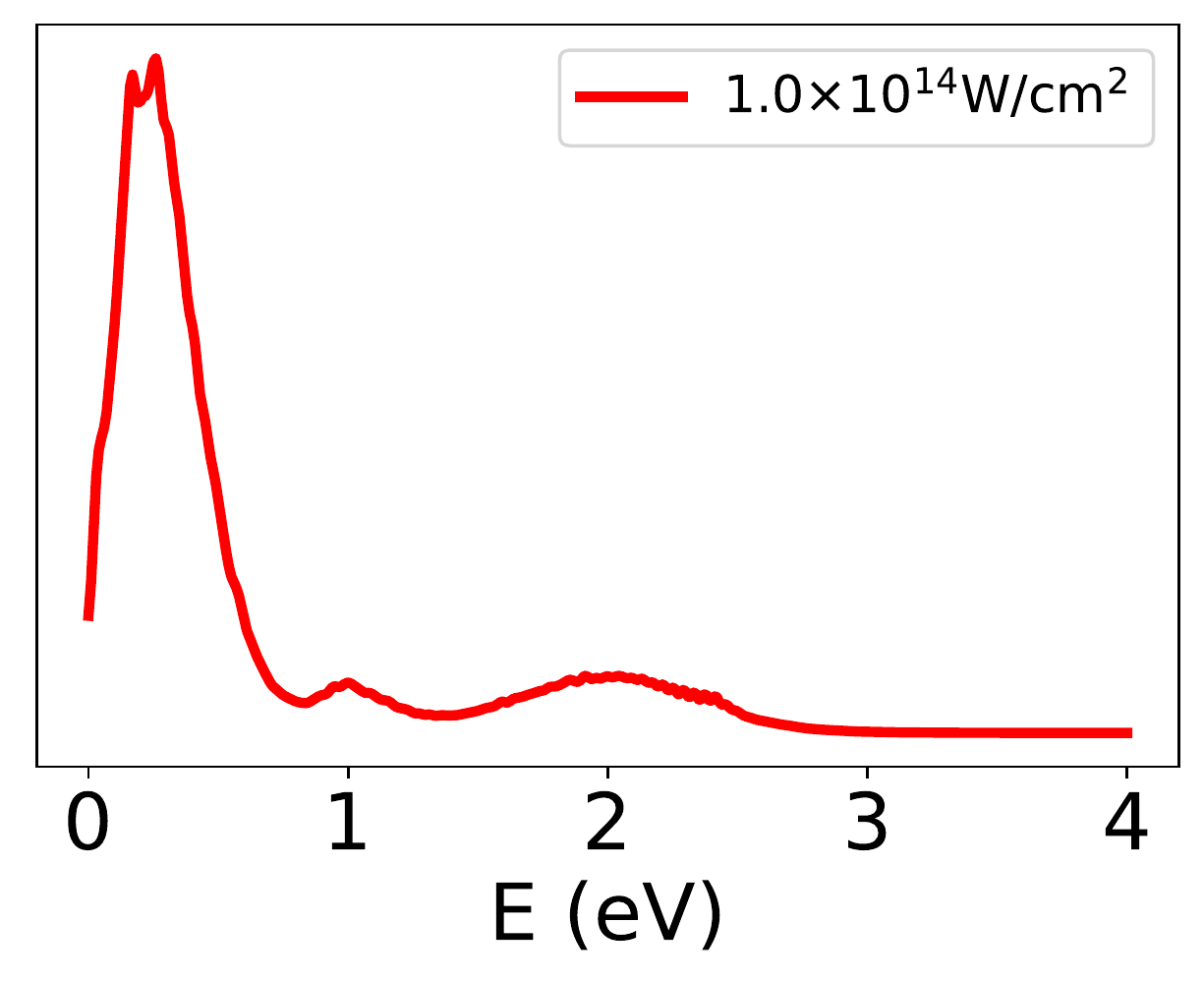}}
 \hfill	
  \subfloat[]{\includegraphics[height=4cm]{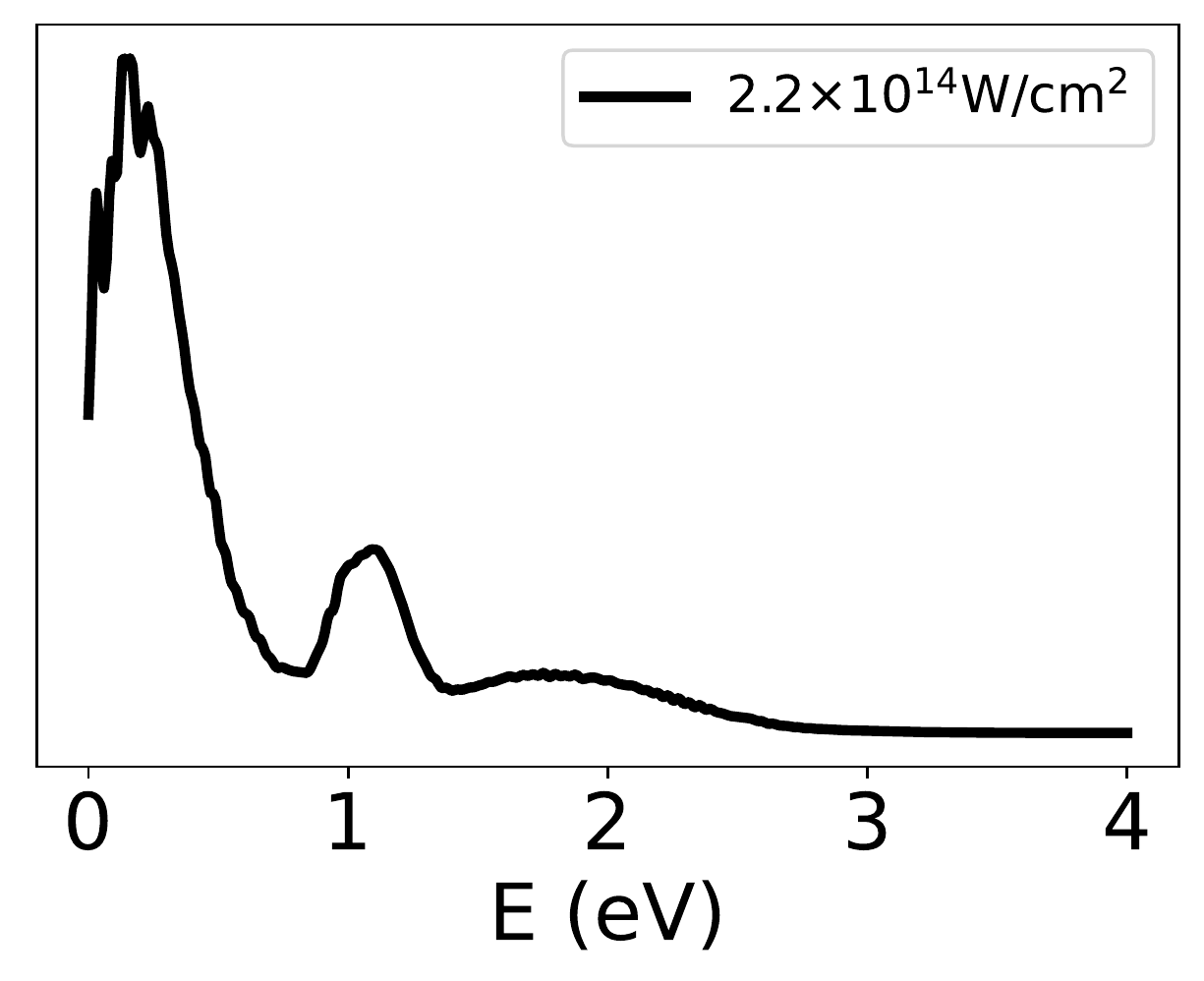}}
  \newline
  \subfloat[]{\includegraphics[height=4.12cm]{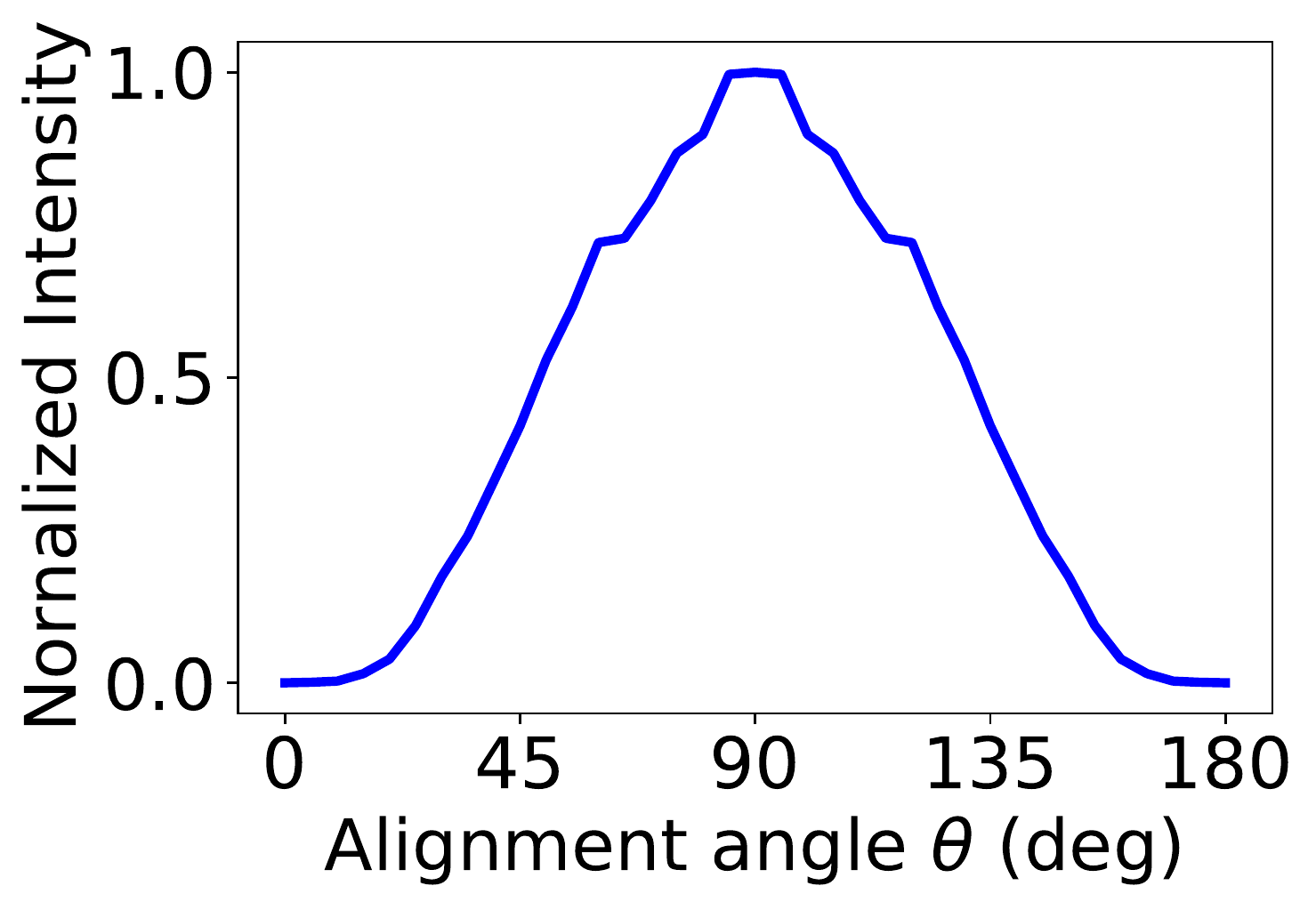}}
 \hfill 	
  \subfloat[]{\includegraphics[height=4cm]{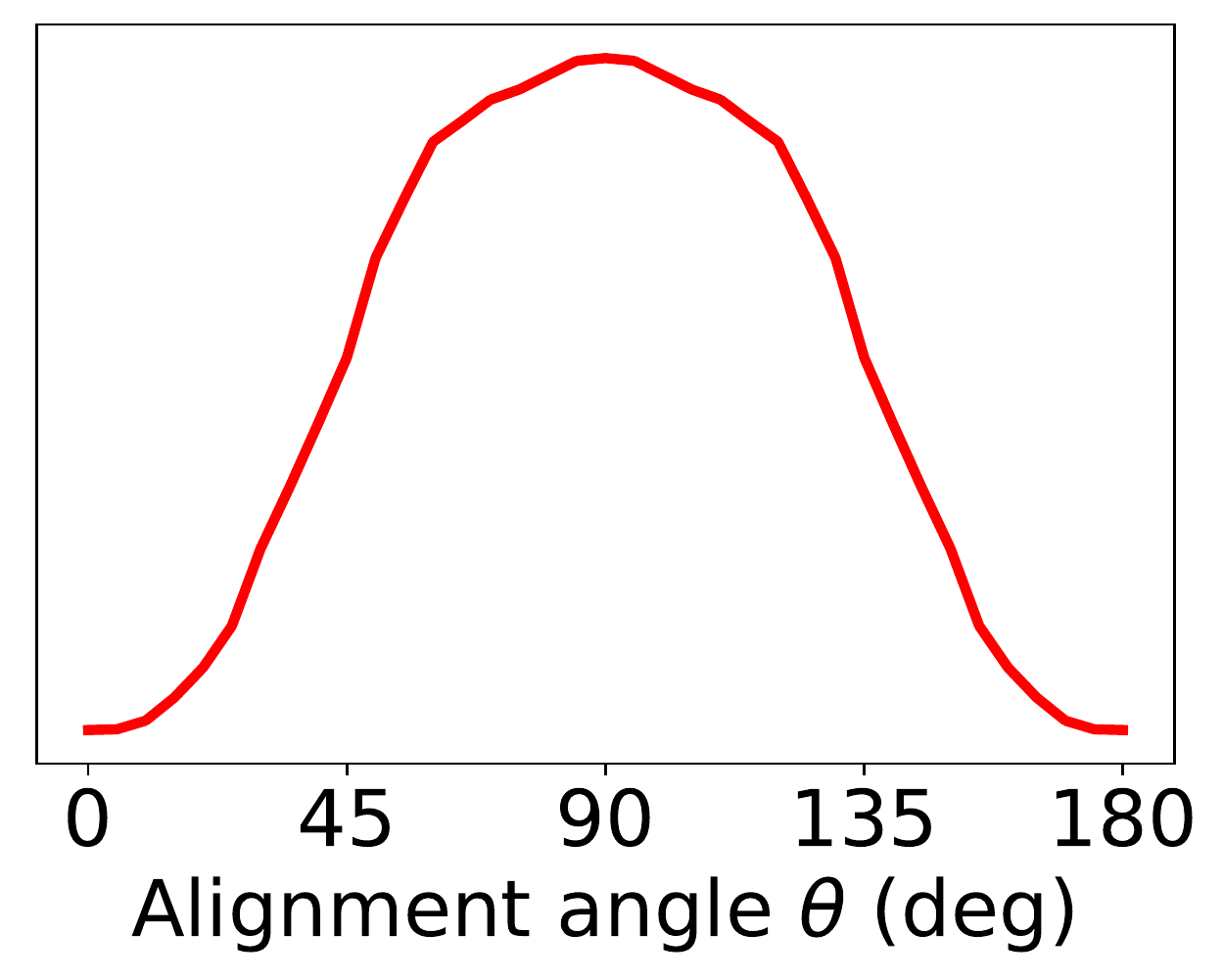}}
 \hfill	
  \subfloat[]{\includegraphics[height=4cm]{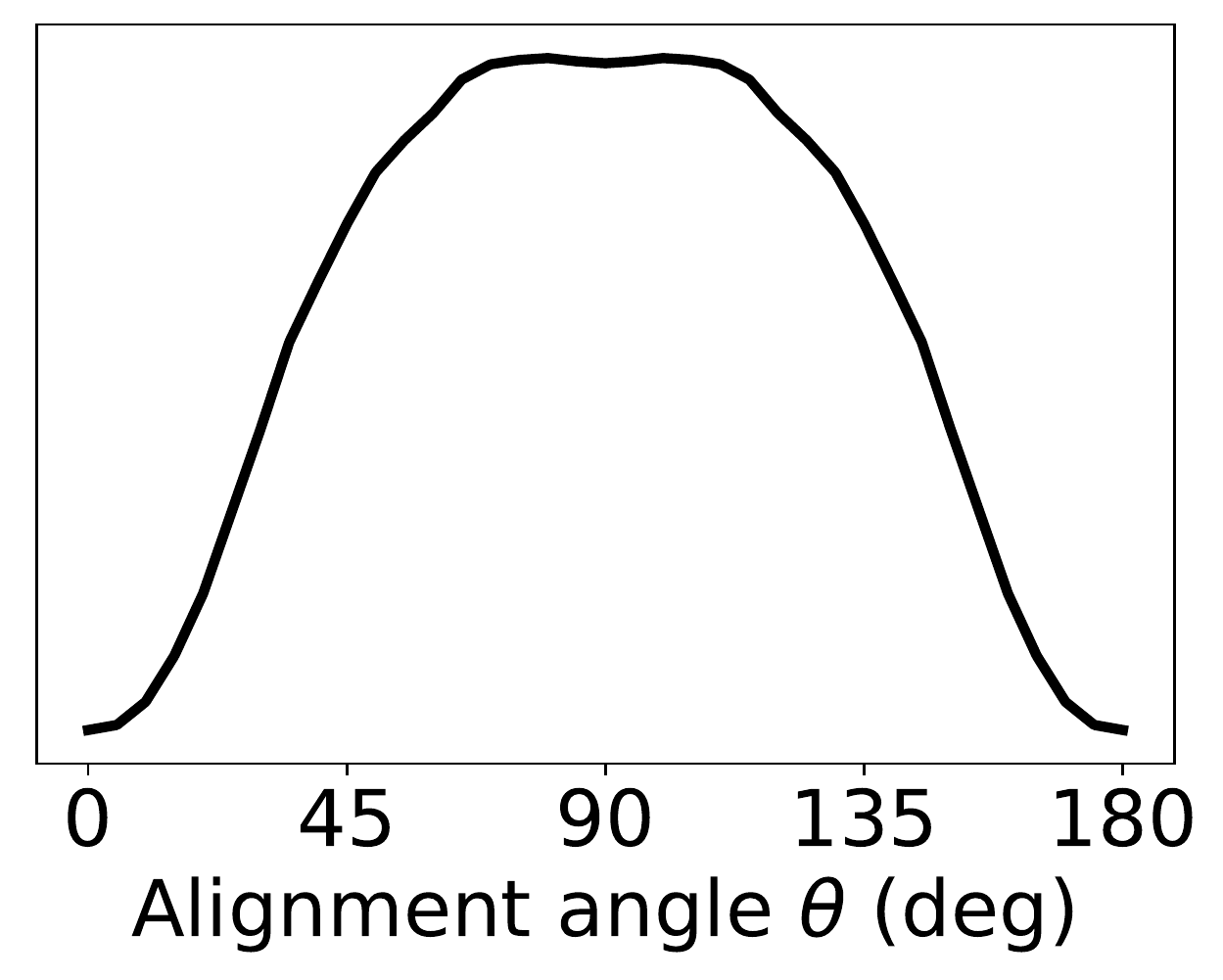}}
    \caption{
    Alignment angle-integrated energy distribution (upper panel) and energy-integrated alignment angle dependence (lower panel) of normalized dissociation yields $D(\theta,E)$ of N+N$^+$ fragments with kinetic energies from 0 eV to 0.8 eV.
    From left to right columns, the peak intensity of laser is $0.5\times 10^{14}$  W/cm$^2$, $1.0\times 10^{14}$ W/cm$^2$ and $2.2\times 10^{14}$ W/cm$^2$, respectively.}
    \label{fig:ker1}
\end{figure}

%%%%%%%%%%%%%%%%%%%%%%%%%%%%%%%%%%%%%
%
In \fig{fig:ker1}, we present the alignment angle-integrated dissociation $D(\theta,E)$ spectra for ionization pulse peak intensities of $0.5\times 10^{14}$ W/cm$^2$, $1.0\times 10^{14}$ W/cm$^2$ and $2.2\times 10^{14}$ W/cm$^2$, respectively.

\begin{figure}[htb!]
  \subfloat[]{\includegraphics[height=4.72cm]{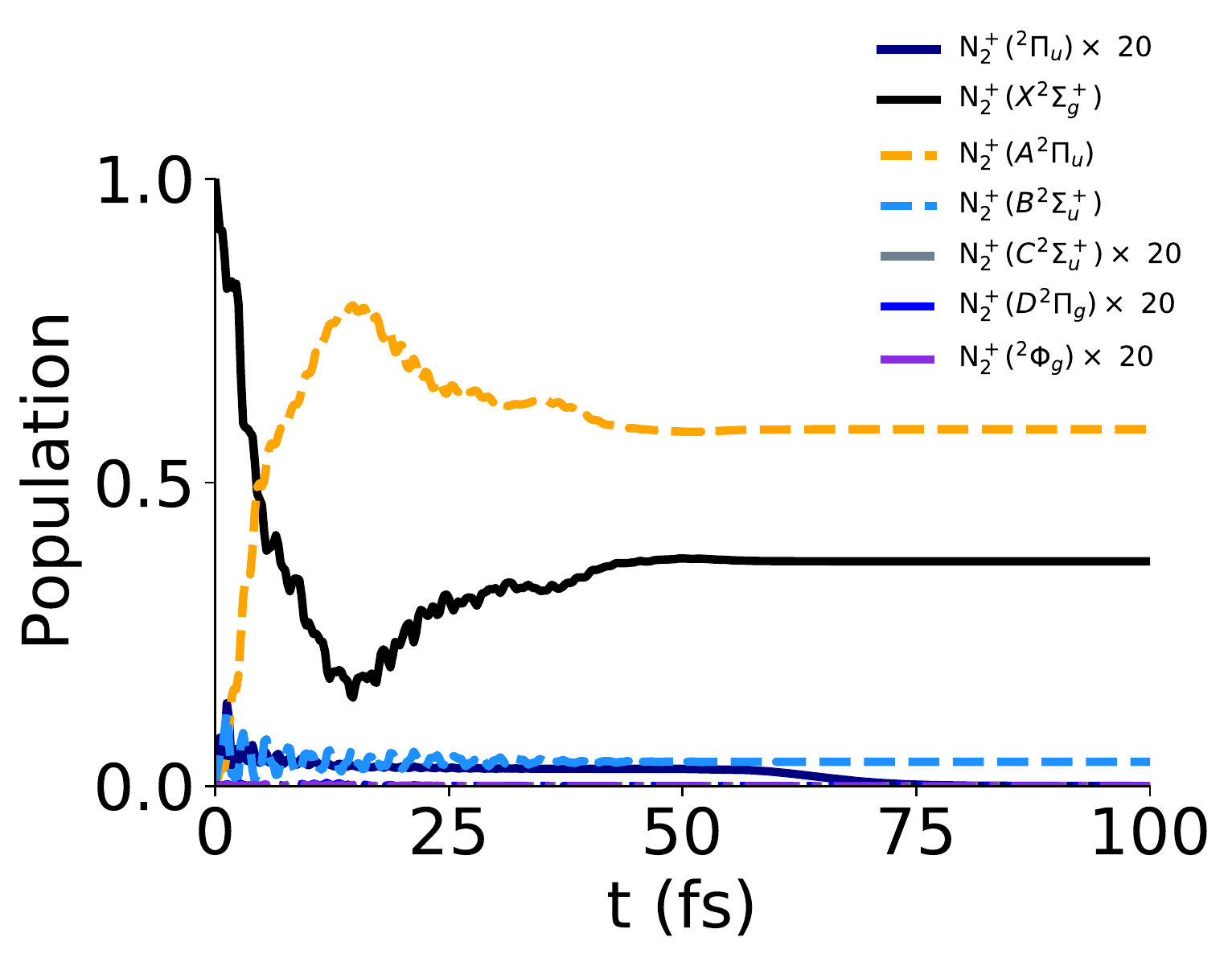}}
 \hfill 	
  \subfloat[]{\includegraphics[height=4.8cm]{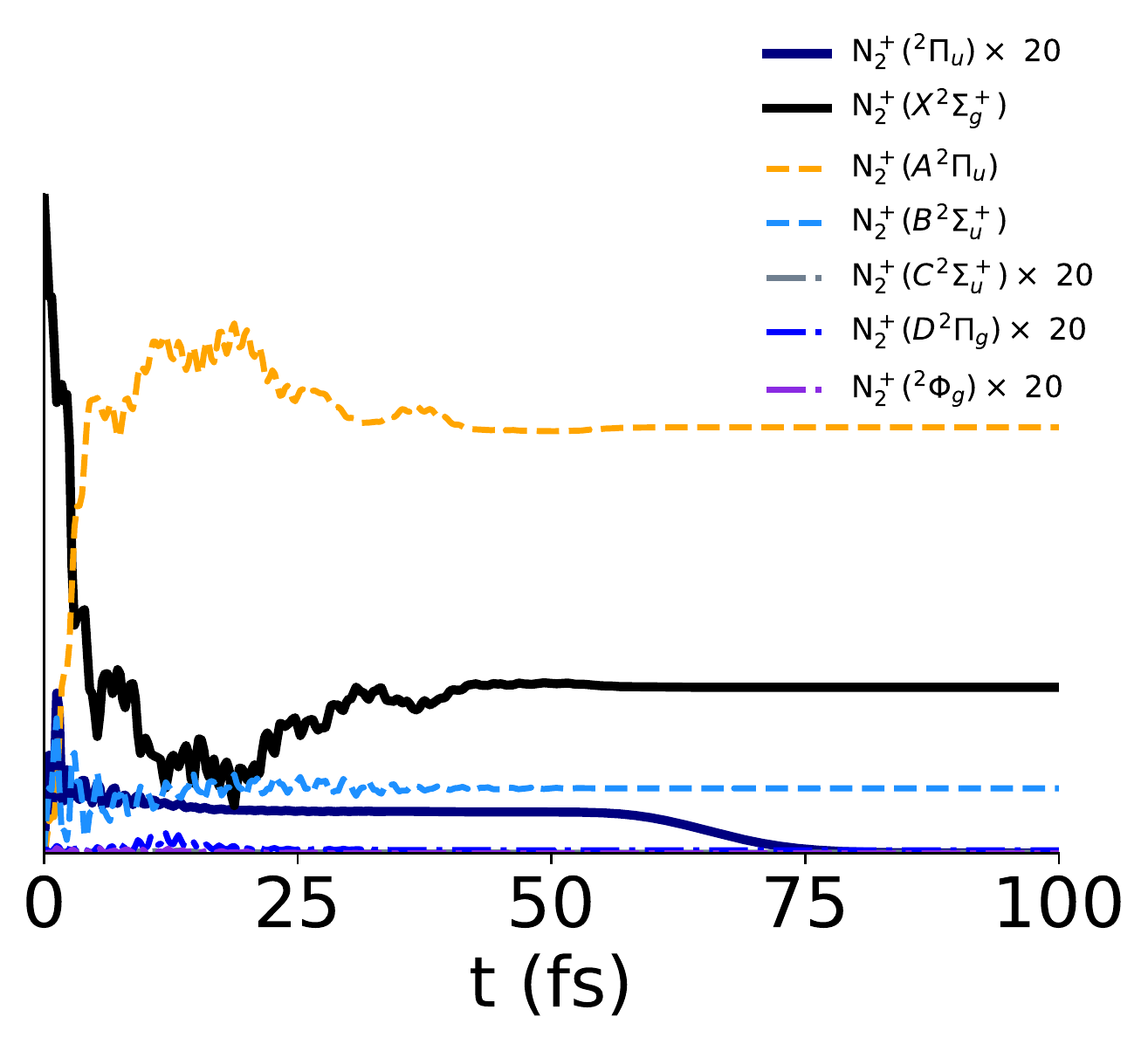}}
 \hfill	
  \subfloat[]{\includegraphics[height=4.8cm]{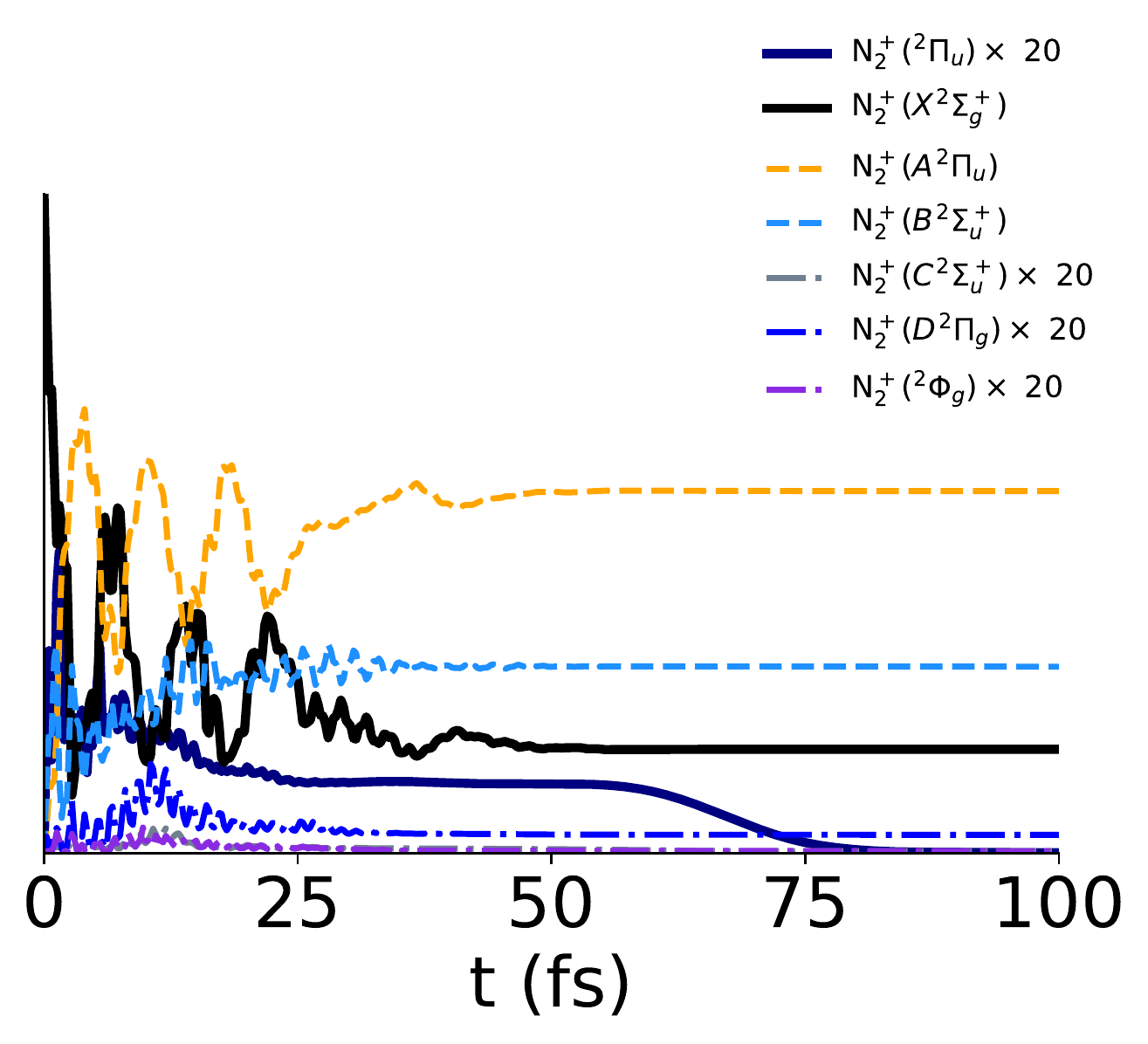}}
    \caption{The temporal population evolution of $\piu$ and intermediate states $\Bsup, \Csup, \Dpig$, $\phg$ of N$_2^+$ at $0.5\times 10^{14}$ W/cm$^2$, $1.0\times 10^{14}$ W/cm$^2$ and $2.2\times 10^{14}$ W/cm$^2$, respectively, and at alignment angle $\theta=60$ degrees.}
    \label{fig:v158_popu}
\end{figure}

For the dissociation dynamics, the contribution from initial states $\Apiu$ and $\Bsup$  is negligible. 
Because in the MO-ADK calculation, even in field intensity of 2.2$\times 10^{14}$W/cm$^2$, ionization rate $W_\nu$ of $\Xsgp$, are two orders of magnitude higher than that of $\Apiu$ and $\Bsup$.
As demonstrated in Eq.~\ref{eq:ker_tot}, the final KER intensity is dependent on both the ionization rate and dissociation rate.
In our MCTDH calculation from equally populated initial condition, the ratio of dissociation yields between $\Xsgp$, $\Apiu$ and $\Bsup$ states are 764: 75: 221, which are of similar magnitude.
However, multiplying the ionization rate, the KER intensity of $\Xsgp$, $\Apiu$, $\Bsup$ initial states are approximately $7.6\times10^{-15}$ a.u, $7.5\times10^{-18}$ a.u and $1.1\times10^{-17}$ a.u, respectively.
For N$_2^+$ and O$_2^+$ cations considered in this work, the single initial state approximation is a reasonable assumption.
In the case that multiple initial states must be taken into account, one has to get the state-resolved dissociation yields from Eq.~\ref{eq:ker_tot},incorporating the calculated ionization rate $W_\nu$ to different initial state.
The energy-integrated and alignment angle-resolved yields reveal that the $D(\theta,E)$ maximizes at $\theta=\pi/2$ and minimizes at $\theta=0$ and $\pi$. 
We assume that at $t=0$ the molecular nitrogen ions are generated from neutral nitrogen molecules and populate the low-lying excited states.
The population transfer starts to occur due to the coupling induced by the remaining laser field.
As shown in \fig{fig:v158_popu}, from 0 to 5 fs most population in $\Init$($\Xsgp$) state is transferred to $\Init$($\Apiu$) state, the population in the $\Init$($\Apiu$) state is greatly enhanced, forming population inversion relative to the ground state $\Xsgp$.
Meanwhile, certain amount of population is transferred to the dissociative $\Init$($\piu$) state. The symmetry selection rule permits first order coupling between $\Init$($\Xsgp$) and $\Init$($\piu$).
In the period between 5 and 10 fs, the population of $\Init$($\Xsgp$) is transferred through the intermediate states ($\Bsup$, $\Csup$, $\Dpig$ and $\phg$) through the higher order pathways as listed in \tab{tab:X2Piu}.
Since the transition dipole moments between $\Xsgp$ and $\piu$, $\Bsup$ and $\Dpig$, $\Bsup$ and $\phg$, $\Csup$ and $\Dpig$, $\Csup$ and $\phg$ are perpendicular to the molecular axis, the projection of the electric field on the molecular axis amounts to an effective field strength $ E_{\textrm{eff}\perp}=E_0 \sin \theta$, where $E_0 \propto \sqrt{I_0}$ is the peak field strength of the probe pulse.
And the transition-dipole moment between $\Xsgp$ and $\Bsup$, $\Xsgp$ and $\Csup$, $\Dpig$ and $\piu$, $\phg$ and $\piu$ is parallel to the molecular axis the projection of the electric field on the molecular axis amounts to an effective field strength $E_{\textrm{eff}\parallel}=E_0 \cos \theta$.
Thus the angular dependence of high order transitions in $E_{\textrm{eff}\perp}$ and in $E_{\textrm{eff}\parallel}$ translate into a spectrum structure between 0 and $\pi/2$, $\pi/2$ and $\pi$.
The analysis above suggests that the intensity dependence of alignment angle-resolved yields spectra contains information about the population dynamics and can be used to quantitatively analyze the pathways of PPRM.

After ca. 50 fs, the population reaches equilibrium because the laser coupling disappear.
The simulation shows that the N$_2^+$ dissociation is dominated by the $\piu$ state, the wavepacket reaches the flux plane at ca. 70 fs. The $^2\Phi_u$ state also provides minor contribution to dissociation, which does not exceed 10\% at highest pulse intensity in our study.

Next we examine the interpretation of alignment angle-resolved yields based on PPRM for various ionization pulse intensities.
As shown in \fig{fig:ker1} (d)(e)(f), the alignment angle-resolved yields exhibit apparent broadening as the peak intensity of ionization pulse increases.

\begin{figure}[htb!]
    \centering
    \includegraphics[width=0.5\textwidth]{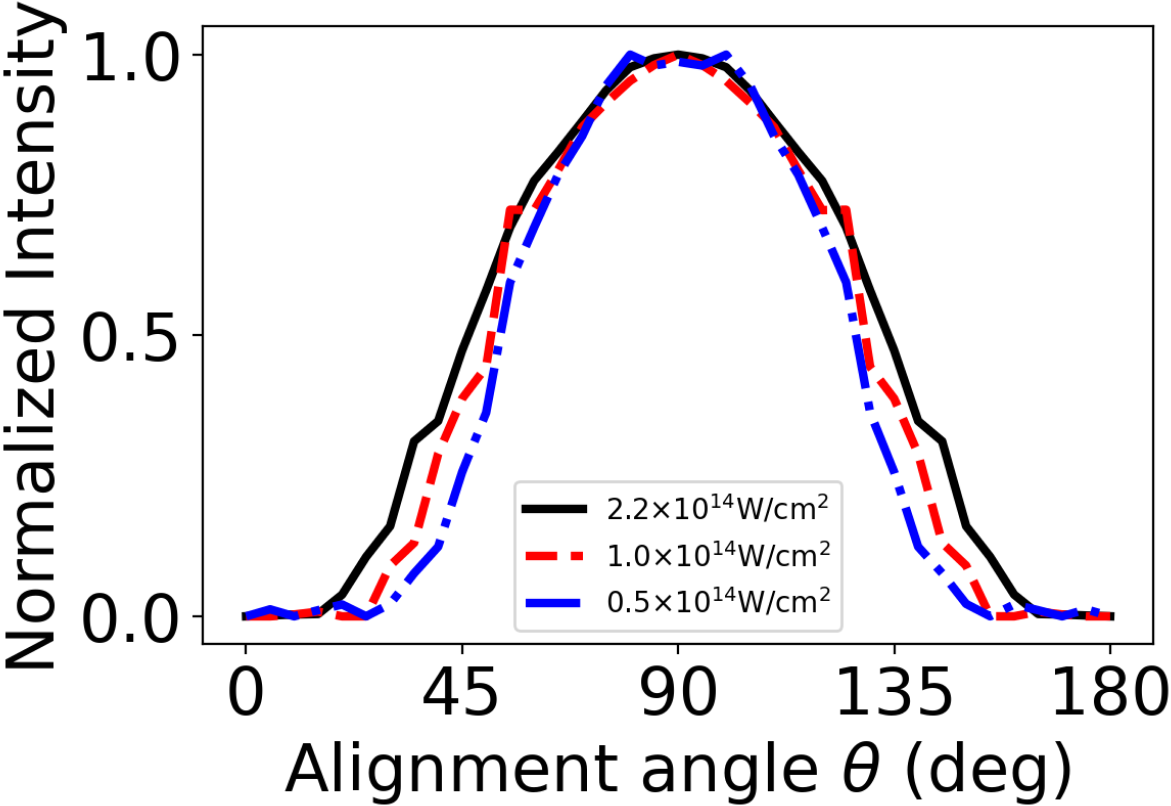}
    \caption{The angular dependence of dissociation yields when only first order pathway between $\Xsgp$ and $\piu$ states exists. From left to right columns, the intensity of laser is $0.5\times 10^{14}$ W/cm$^2$, $1.0\times 10^{14}$ W/cm$^2$ and $2.2\times 10^{14}$ W/cm$^2$, respectively.}
    \label{fig:ker2}
\end{figure}

It should be noted that although there is direct coupling between the $\Init$($\Xsgp$) and dissociative $\piu$ state, the participation of the intermediate states $\Init$($\Bsup$, $\Csup$, $\Dpig$ and $\phg$) is indispensable for generating the observed intensity dependent angle distribution at the non-perpendicular angle.
As evidenced by \fig{fig:ker2}, if we remove the coupling between $\Init$($\Xsgp$) and intermediate states($\Bsup$, $\Csup$, $\Dpig$ and $\phg$), dissociated states $\piu$ and those intermediate states, the dissociation will dominantly take place for $\theta=\pi/2$ and exhibit almost no intensity dependence.

%%%%%%%%%%%%%%%%%%%%%%%%%%%%%%%%%%%%%%%%%
%%%%%%%%%%%%%%%%%%%%%%%%%%%%%%%%%%%%%%%%%

%%%%%%%%%%%%%%%%%%%%%%%%%%%%%%%%%%%%%%%%%

\begin{table}[htb!]
    \centering
    \begin{tabular}{|c|c|}
    \hline
    Channel & $(\theta, E)$-dependence \\
    \hline\hline $\piu\muxyarrow\Xsgp$ & $E\sin\theta$ \\
    \hline $\piu\muzarrow\Dpig\muxyarrow\Bsup\muzarrow\Xsgp$ & $E^3\cos^2\theta\sin\theta$ \\
    $\piu\muzarrow\ \phg\muxyarrow\Bsup\muzarrow\Xsgp$ &  $E^3\cos^2\theta\sin\theta$ \\
    $\piu\muzarrow\Dpig\muxyarrow\Csup\muzarrow\Xsgp$ &  $E^3\cos^2\theta\sin\theta$ \\
    $\piu\muzarrow\ \phg\muxyarrow\Csup\muzarrow\Xsgp$ &  $E^3\cos^2\theta\sin\theta$\\
    \hline
    \end{tabular}
    \caption{The first and third order pathways from $\Xsgp$ to $\piu$ and the dependence of transition amplitude on electric field strength and alignment angle. $\muxy$ denotes transition dipole along perpendicular to the molecular axis and $\muz$ denotes transition dipole along parallel direction.}
    \label{tab:X2Piu}
\end{table}
%%%%%%%%%%%%%%%%%%%%%%%%%%%%%%%%%%%%%%%%%

%%%%%%%%%%%%%%%%%%%%%%%%%%%%%%%%%%%%%
%%%%%%%%%%%%%%%%%%%%%%%%%%%%%%%%%%%%%%

\subsection{Dissociation dynamics of O$_2^+$ cation}
\begin{figure}[htb!]
    \centering
    \includegraphics[width=0.7\textwidth]{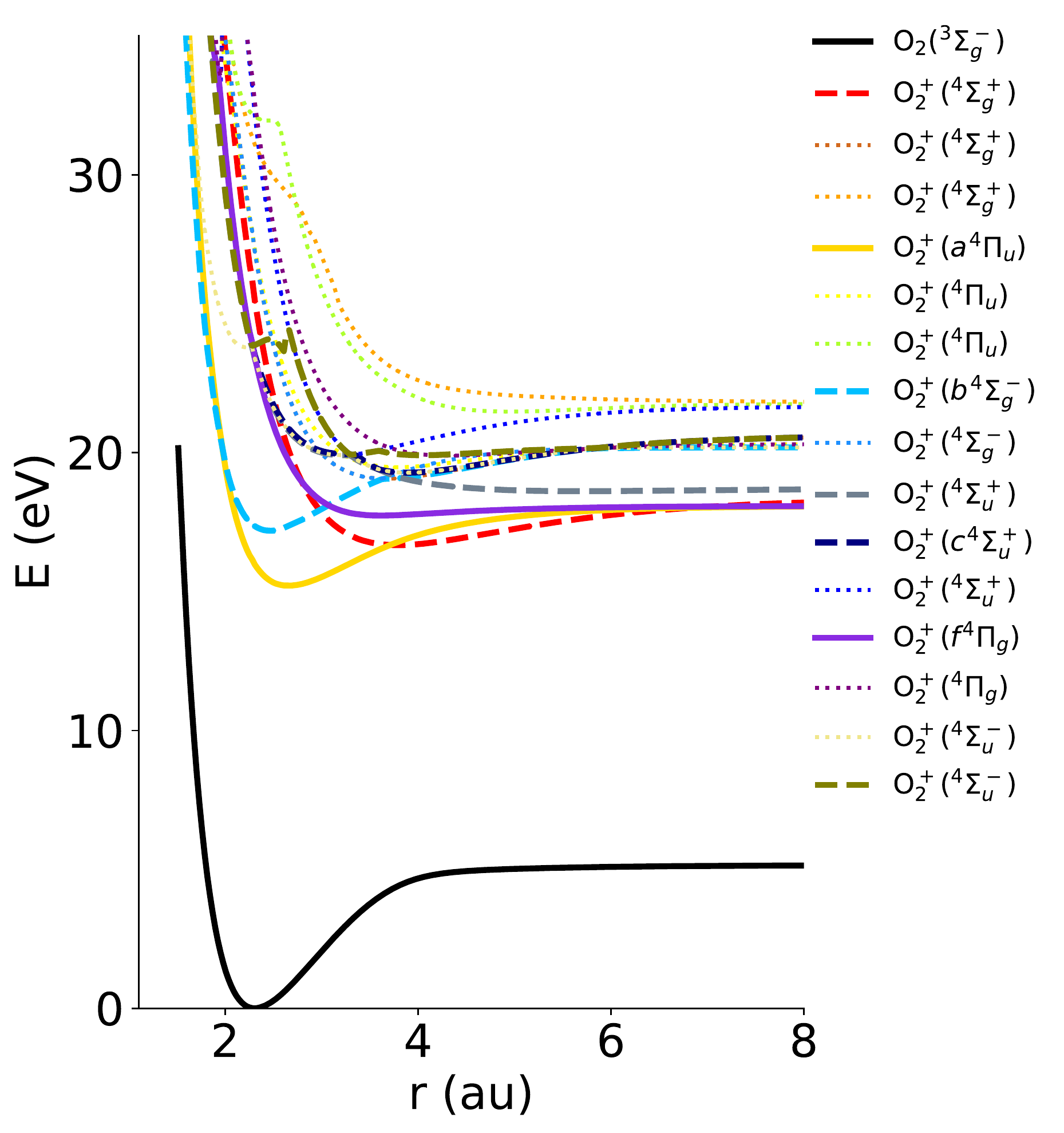}
    \caption{The potential energy curves of all electronic states of O$_2^+$ cation. 
    Yellow solid line is initial exicted state $\apiu$.
    Blue solid line is final dissociated state $\fpig$.
    Light blue, dark blue, green, gray, and red dashed lines are for the intermediate state $\bsgm$, $\csup$, $\dsum$,  $\esum$ and $\hsgp$, respectively, which actively participate in the population transfer dynamics.
    }
    \label{fig:o2pec}
\end{figure}

\begin{figure}[htb!]	 	
  \subfloat[]{\includegraphics[height=4.12cm]{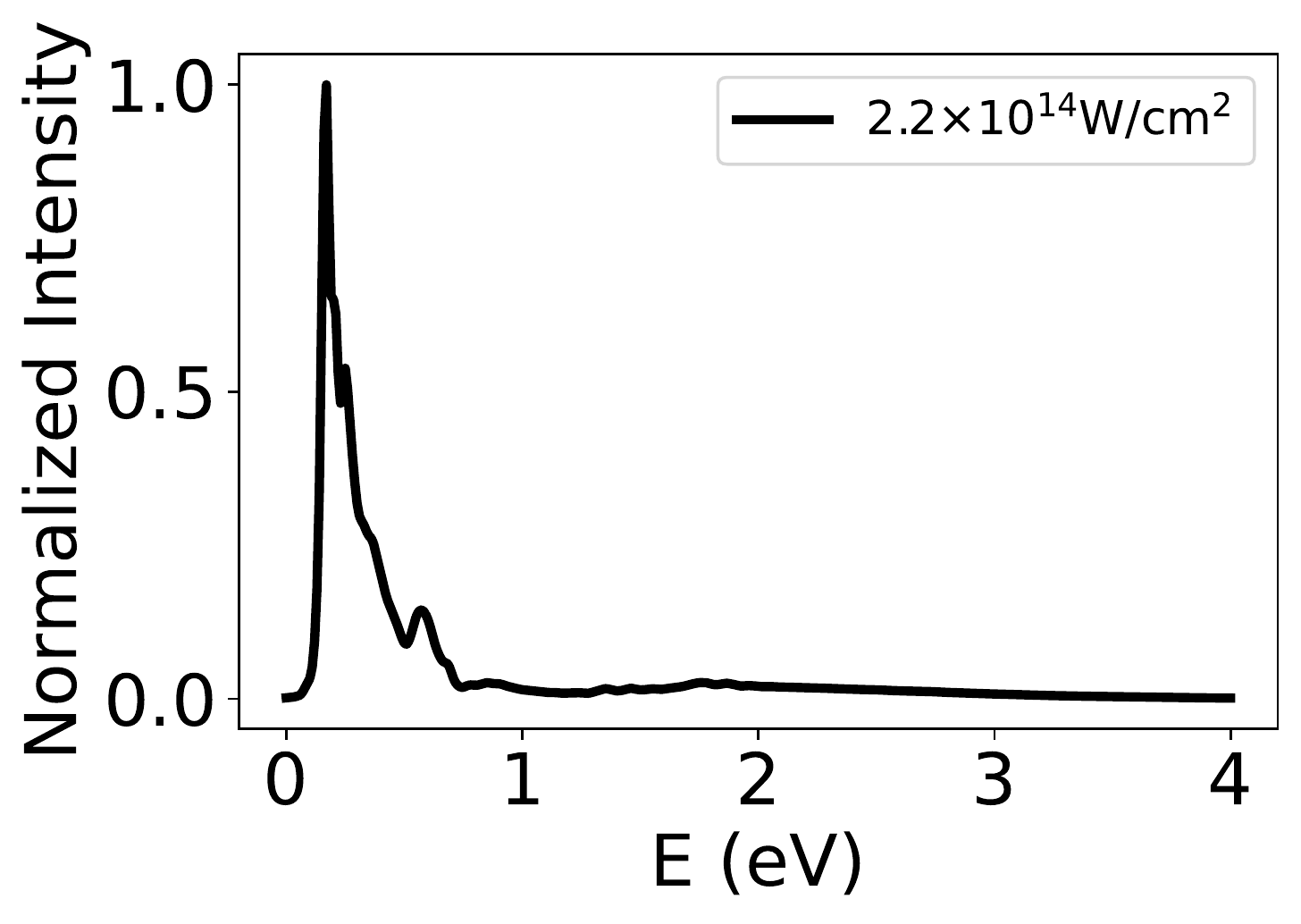}}
 \hfill 	
  \subfloat[]{\includegraphics[height=4cm]{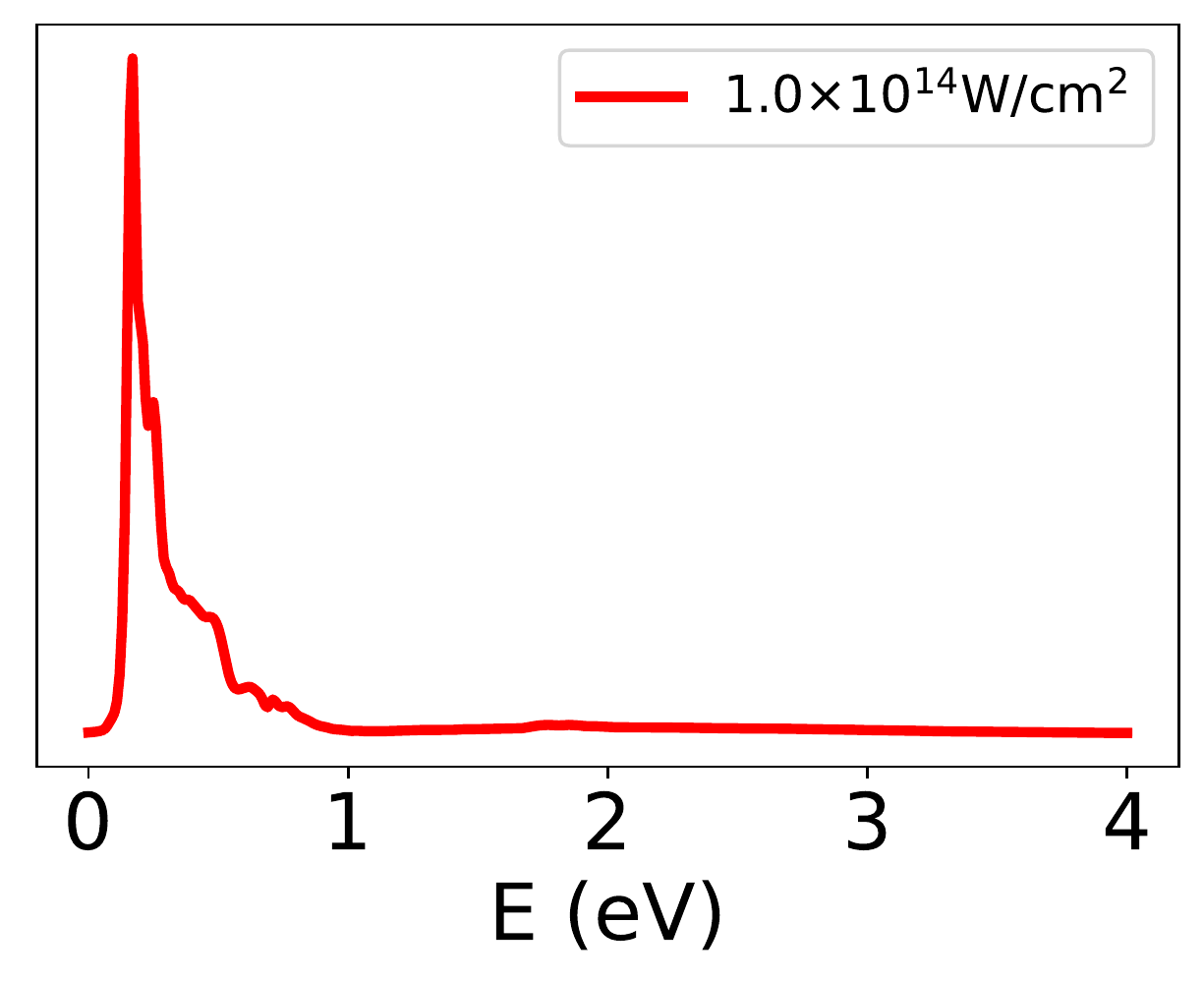}}
 \hfill	
  \subfloat[]{\includegraphics[height=4cm]{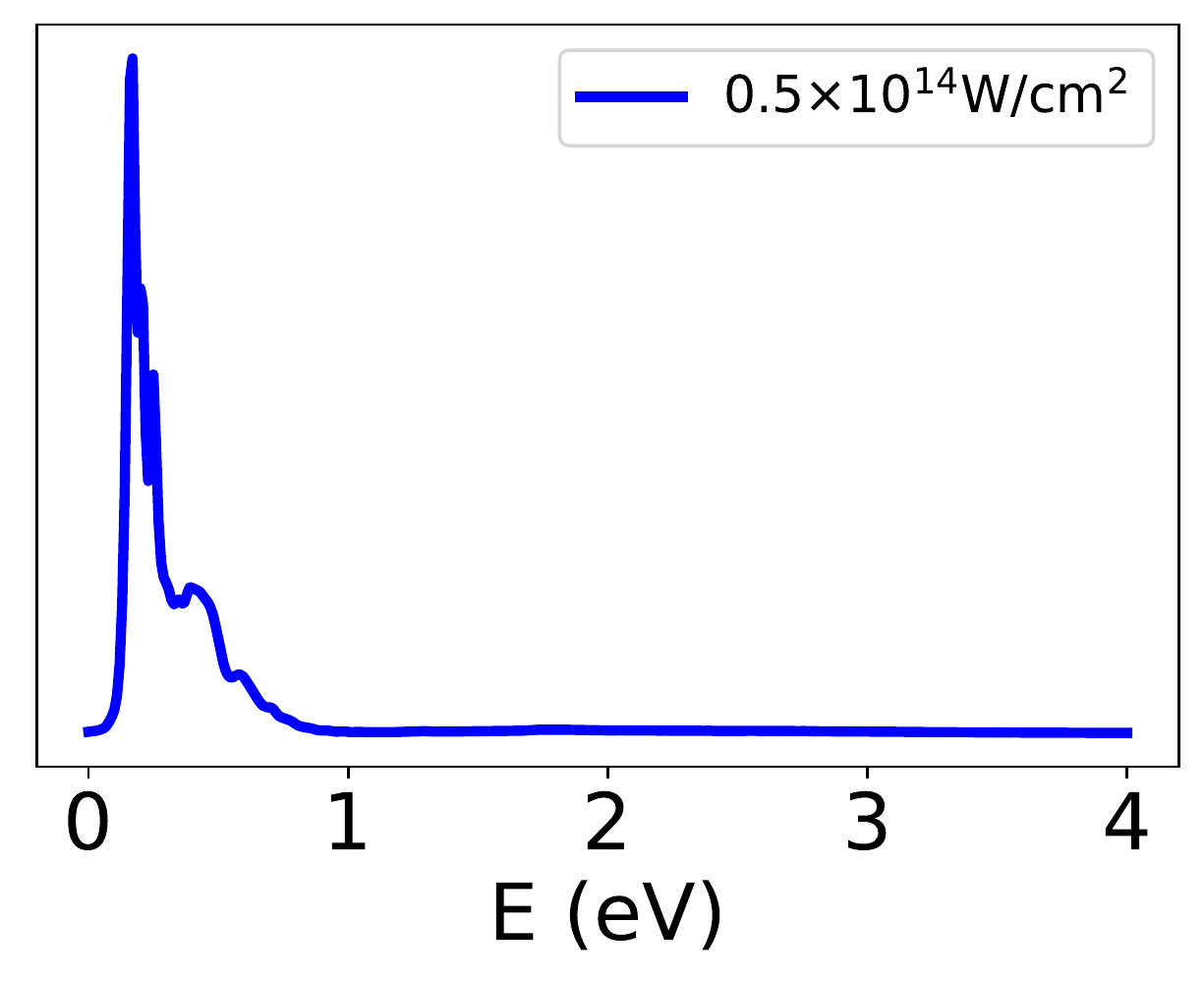}}
  \newline
  \subfloat[]{\includegraphics[height=4.12cm]{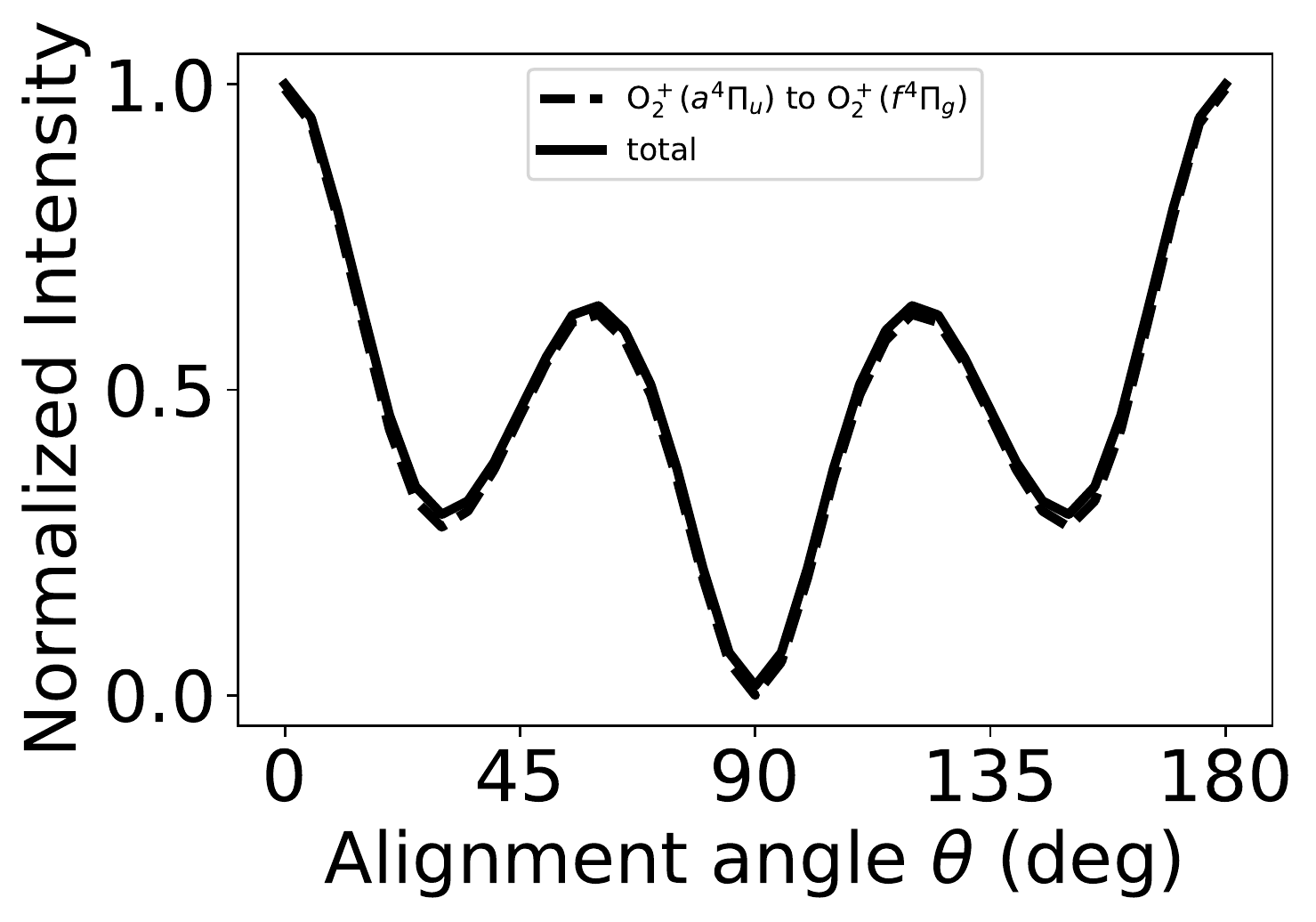}}
 \hfill 	
  \subfloat[]{\includegraphics[height=4cm]{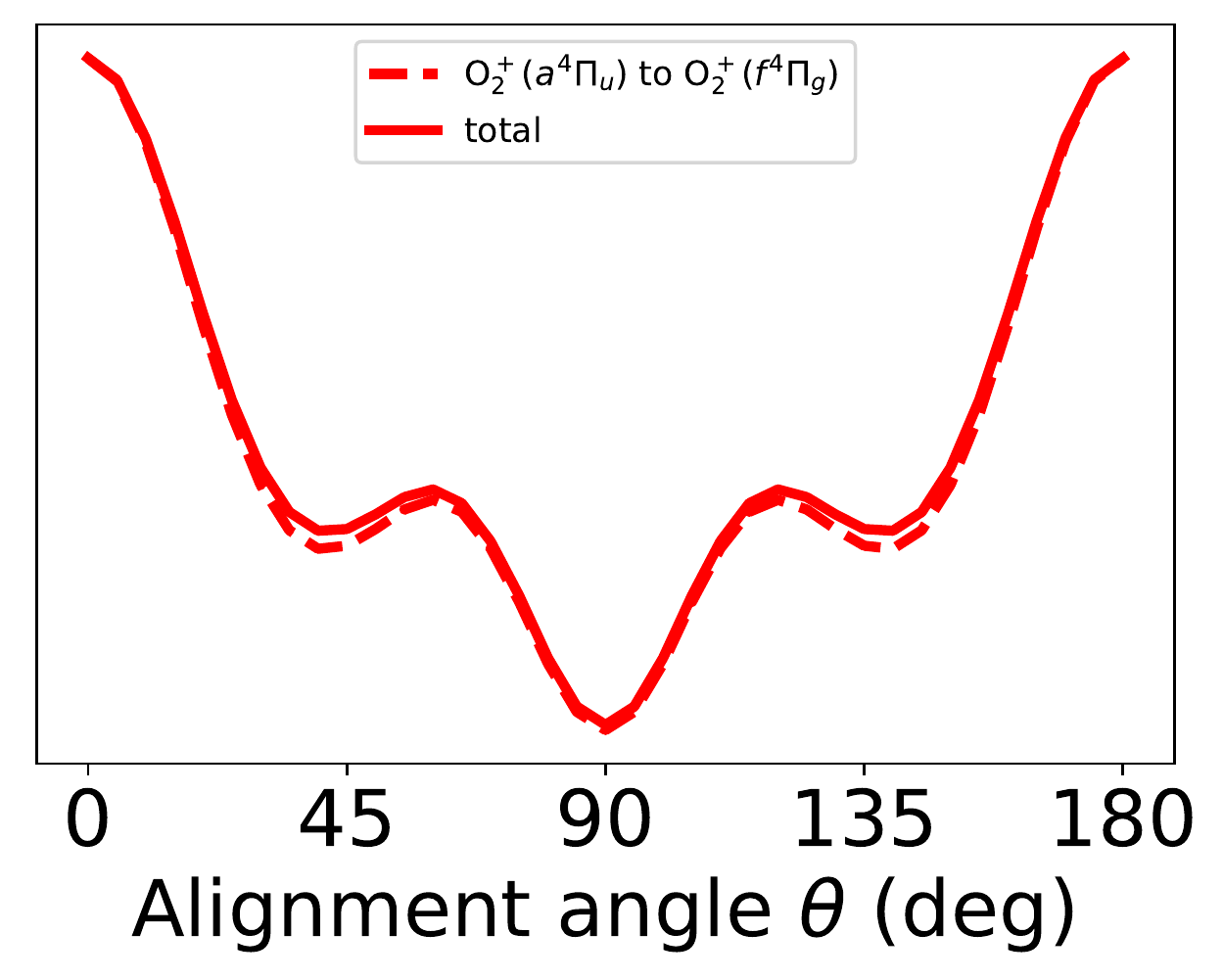}}
 \hfill	
  \subfloat[]{\includegraphics[height=4cm]{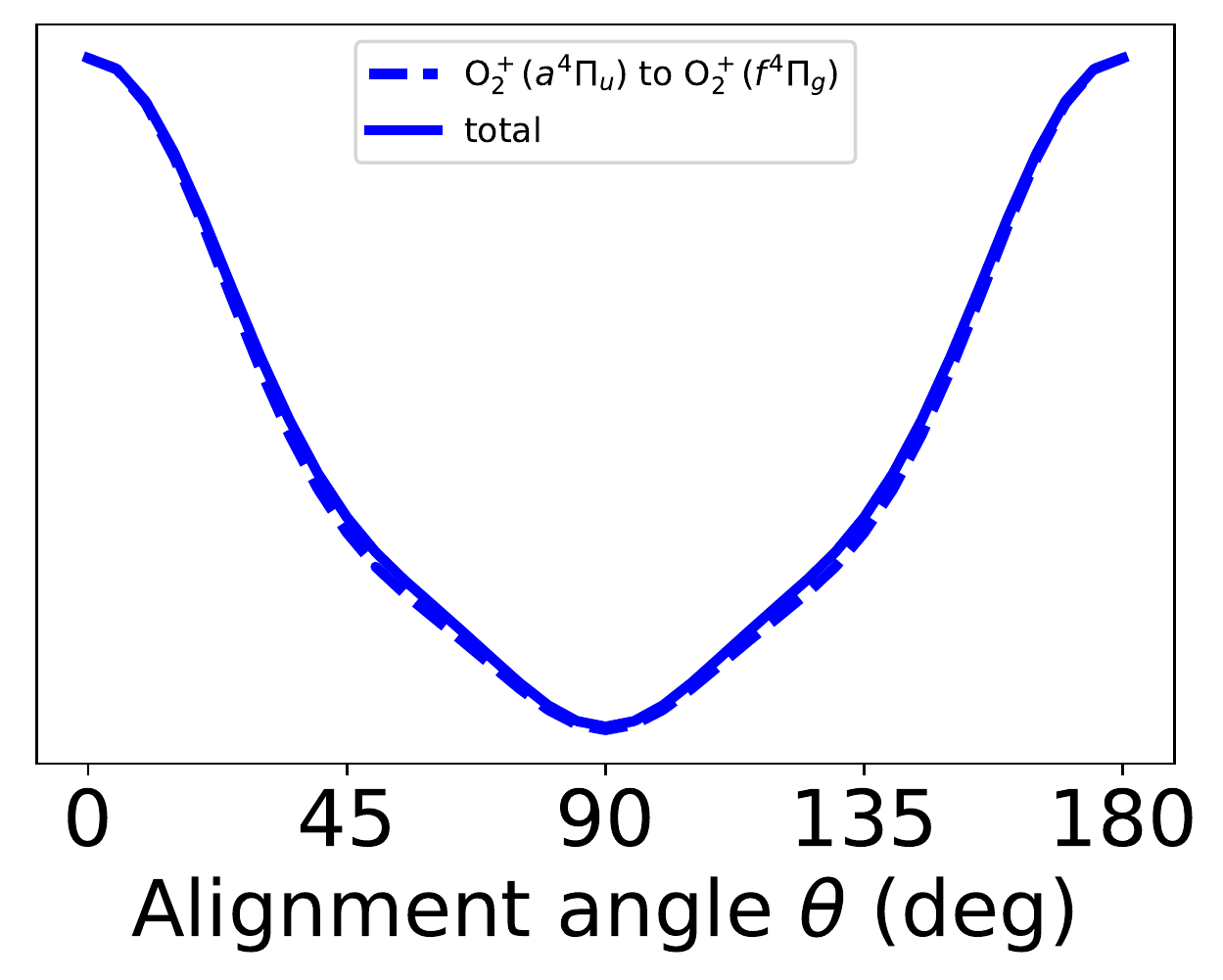}}
    \caption{
    Kinetic energy and alignment angle dependence of O$_2^+$ dissociation induced by strong field ionization.
    The upper panel (a)-(c) are alignment-integrated energy distribution. 
    The lower panel (d)-(f) are energy-integrated angular distribution.
    Dashed lines in (d)-(f) stand for the dissociation yields of the final state $\fpig$.
    From left to right columns, the intensity of laser is $2.2\times 10^{14}$ W/cm$^2$, $1.0\times 10^{14}$ W/cm$^2$ and $0.5\times 10^{14}$ W/cm$^2$, respectively.}
    \label{fig:o2ker1}
\end{figure}

Similarly, the potential energy curves of neutral and cationic oxygen molecule are shown in \fig{fig:o2pec}, and $\apiu$ state is dominantly populated when O$_2$ is ionized via strong field tunneling mechanism.
%%%%%%
In \fig{fig:o2ker1}, we present the normalized angle-integrated yields spectra for ionization pulse with peak intensities of $0.5\times 10^{14}$ W/cm$^2$, $1.0\times 10^{14}$ W/cm$^2$ and $2.2\times 10^{14}$ W/cm$^2$, respectively.
The peak at $\sim$ 0.17 eV is solely contributed from dissociation in $\fpig$ state, which is excited from $\apiu$.
Moreover, the kinetic energy-integrated angle-resolved yields are presented in \fig{fig:o2ker1}.
The laser intensity dependent angular distribution of dissociation yields of O$_2^+$ can be similarly explained with the different pathways of PPRM.
The O$_2^+$ dissociation is dominantly taking place on the $f^4\Pi_g$ state, when the initial wavepacket is launched in the $a^4\Pi_u$ state by strong field ionization~\cite{corlin15:043415,abanador20:043410}.

In the MCTDH simulation, we assume that the cation is generated from neutral oxygen molecules at $t=0$ and dominantly populates the excited state O$_2^+$ ($\apiu$).
During the time period between 0 and 15 fs, most population in O$_2^+$($\apiu$) state is transferred to O$_2^+$($\bsgm$) state.
In the period between 15 and 50 fs, the population of O$_2^+$($\apiu$) is transferred to the dissociative state O$_2^+$($\fpig$) via first order laser-induced coupling and via intermediate states $\bsgm$, $\csup$, $\dsum$, $\esum$ and $\hsgp$ due to the high-order laser-induced coupling, as listed in \tab{tab:a2f}.
Since the transition-dipole moments between $\apiu$ and $\hsgp$, $\apiu$ and $\bsgm$, $\esum$ and $\fpig$, $\csup$ and $\fpig$, $\dsum$ and $\fpig$ are perpendicular to the molecular axis, the projection of the electric field on the molecular axis amounts to an effective field strength $ E_{\textrm{eff}\perp}=E_0 \sin \theta$.
The transition dipole moments between $\apiu$ and $\fpig$, $\hsgp$ and $\esum$, $\hsgp$ and $\csup$, $\bsgm$ and $\dsum$ are parallel to the molecular axis, the projection of the electric field on the molecular axis amounts to an effective field strength $E_{\textrm{eff}\parallel}=E_0 \cos \theta$.
The $\theta$-dependence of high order transition in $E_{\textrm{eff}\perp}$ and in $E_{\textrm{eff}\parallel}$ translates into variation of angular distribution of fragments between $\theta=0$ and $\pi$ as a function of laser intensity.
This fact suggests that the angle distribution of alignment angle-resolved yields contains information about the pathways to reach the dissociative final states, and from which we could classify the pathways of population transfer by their orders.
For time after 50 fs, the population of mainly populated dissociative state $\fpig$ slowly decrease as the wavepacket reaches the flux plane, and in the calculation, the $\fpig$ state contributes to over 97\% of the total dissociation yield.

\begin{figure}[htb!]
    \centering
    \includegraphics[width=0.5\textwidth]{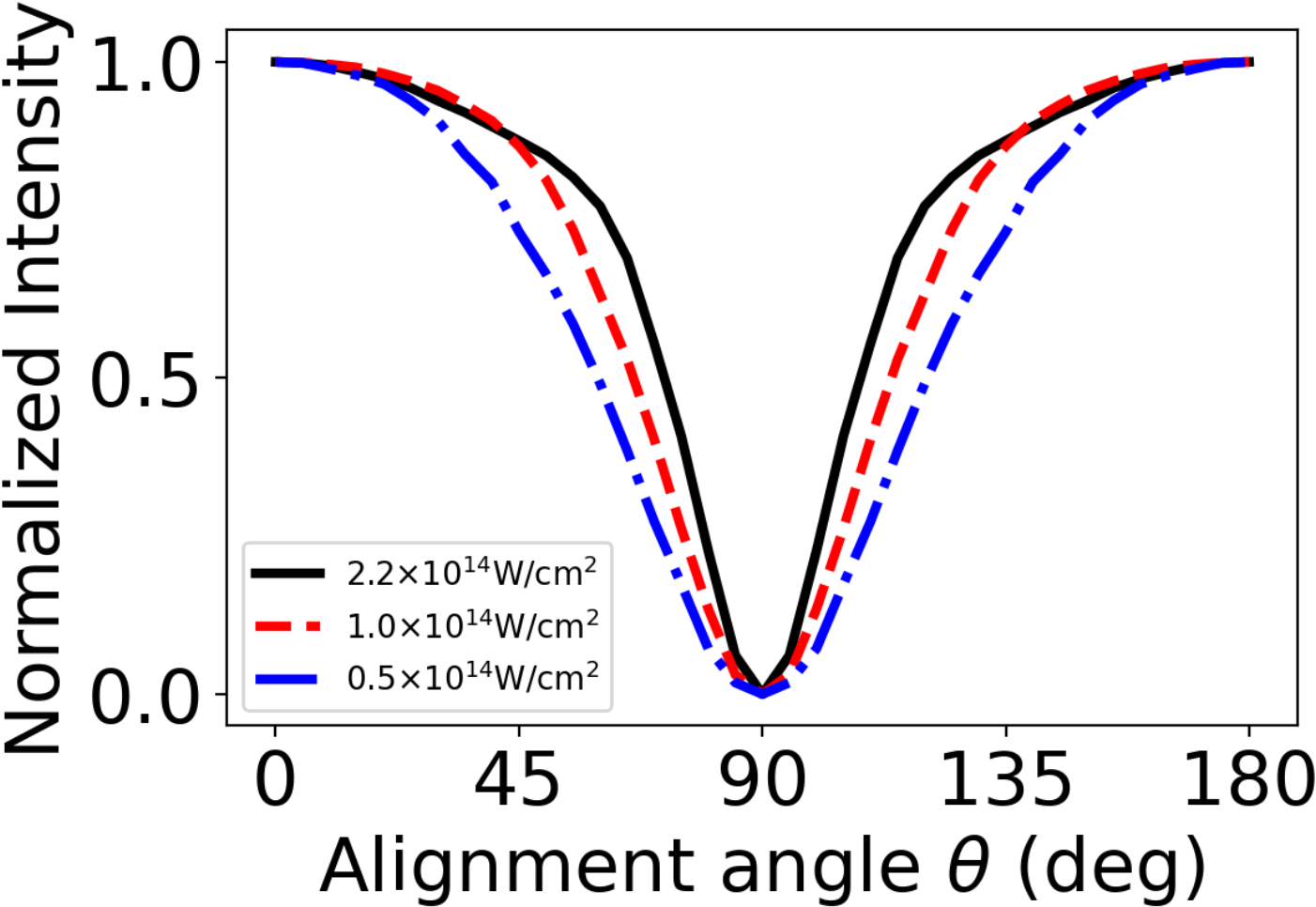}
    \caption{The alignment angle dependence of dissociation yields under various laser intensities, assuming that only first order pathway is present. Black solid line, red dashed line and blue dot dashed line denote the distribution at laser intensity of $2.2\times 10^{14}$ W/cm$^2$, of $1.0\times 10^{14}$ W/cm$^2$ and of $0.5\times 10^{14}$ W/cm$^2$, respectively.}
    \label{fig:o2ker2}
\end{figure}

We then examine the PPRM interpretation in angle-resolve yields spectra for weaker probe-pulse intensities.
For this purpose, we show angle-integrated and energy-integrated yields spectra for probe-pulse peak intensities of $0.5\times 10^{14}$ W/cm$^2$ and $1.0\times 10^{14}$ W/cm$^2$ in \fig{fig:o2ker1}, respectively.
The calculated angle-resolved yields exhibit intensity-dependent angle distribution.
As the peak intensity of ionization laser is increased, the amplitudes of high orders processes become stronger than that of $0.5\times 10^{14}$ W/cm$^2$, and the oscillatory behavior is smeared out in $D(\theta,E)$, as the transitions to intermediate  O$_2^+$ states $\bsgm$, $\csup$, $\dsum$, $\esum$ and $\hsgp$ apparently interfere.
As evidenced by \fig{fig:o2ker2}, if we remove the coupling between O$_2^+$($\apiu$, $\fpig$) and intermediate states ($\bsgm$, $\csup$, $\dsum$, $\esum$ and $\hsgp$), the dissociation will dominantly take place in parallel alignment, and the oscillatory to smooth transition in the alignment angle-dependent yield will not occur.
%%%%%%%
%%%%%%%%%%%%%%%%%%%%%%%%%%%%%%%%%%
%%%%%%%%%%%%%%%%%%%%%%%%%%%%%%%%%%%%%
%%%%%%%%%%%%%%%%%%%%%%%%%%%%%%%%%%%%%%%%%
%%%%%%%%%%%%%%%%%%%%%%%%%%%%%%%%%%%%%%%%%
%%%%%%%%%%%%%%%%%%%%%%%%%%%%%%%%%%%%%%
\begin{figure}[htb!]
  \subfloat[]{\includegraphics[height=4.52cm]{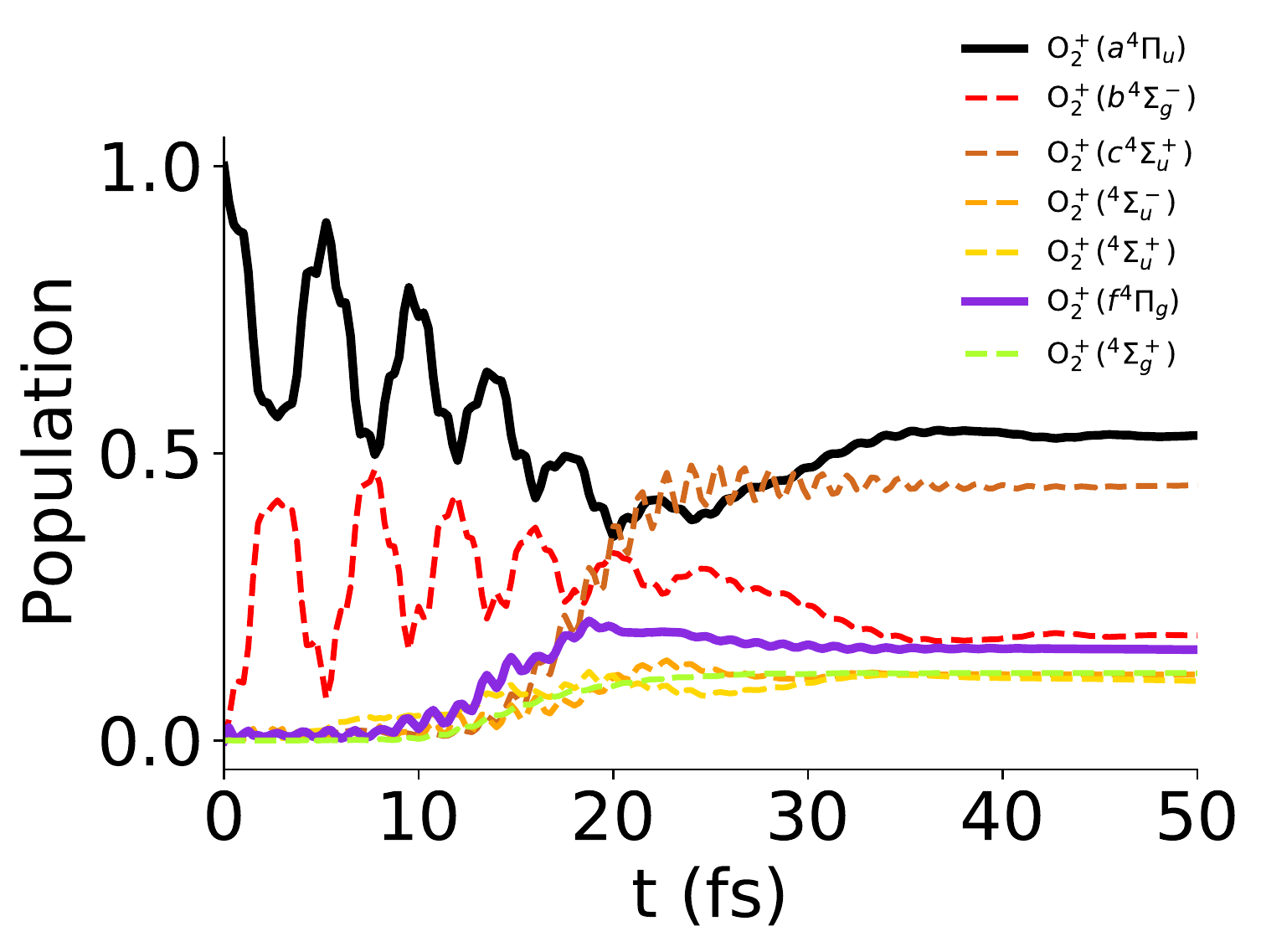}}
 \hfill 	
  \subfloat[]{\includegraphics[height=4.6cm]{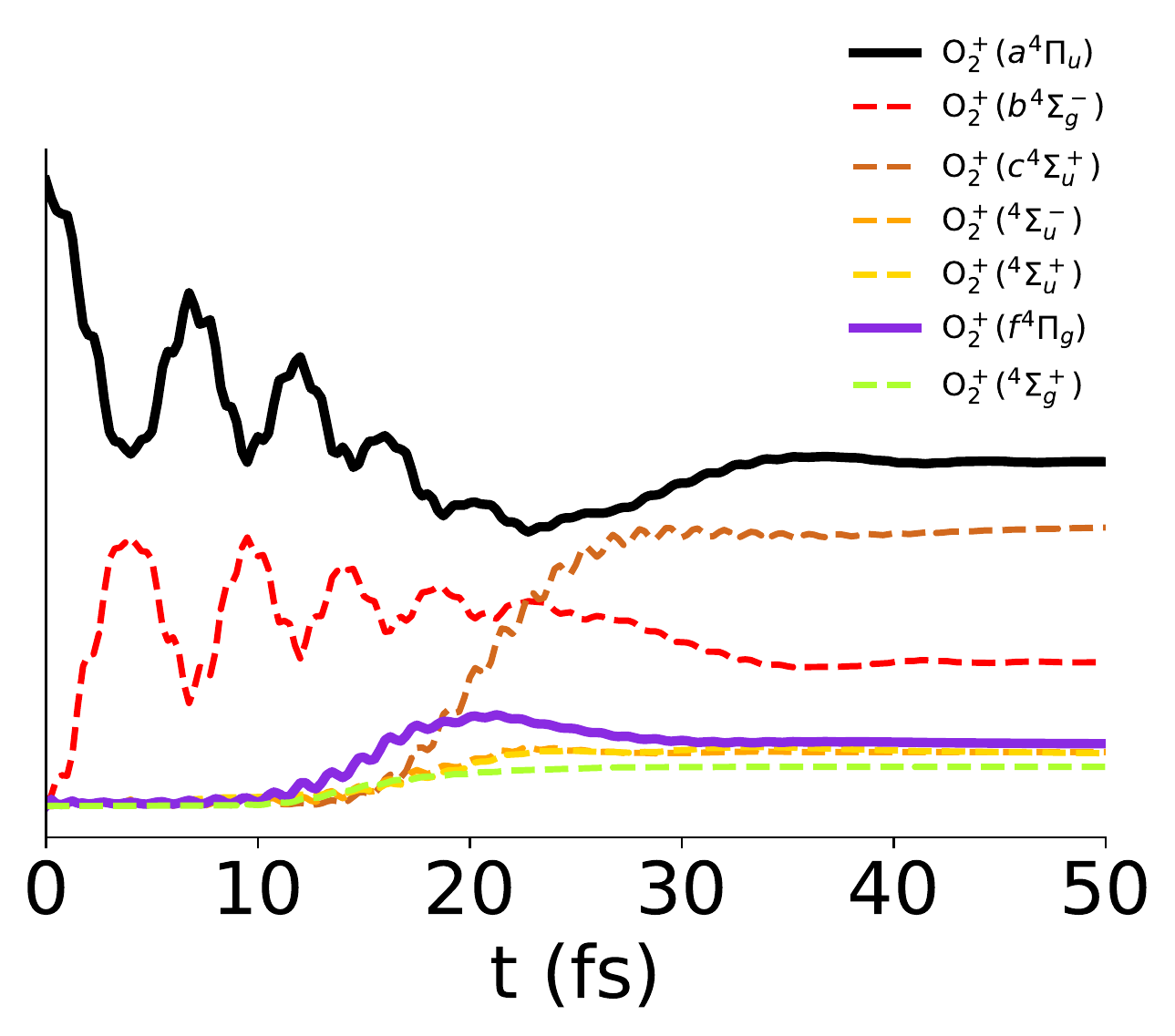}}
 \hfill	
  \subfloat[]{\includegraphics[height=4.6cm]{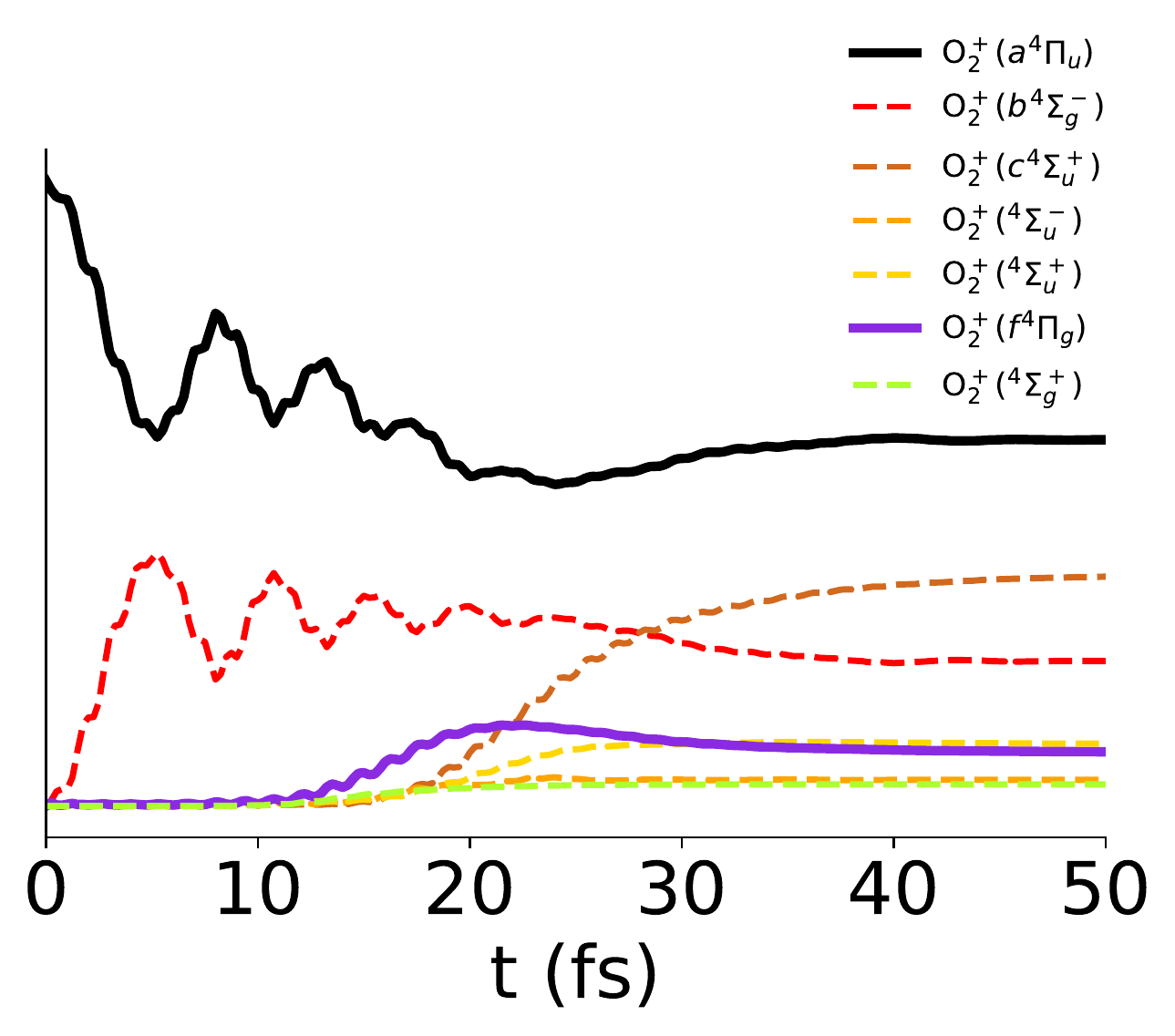}}
    % \centering
    % \includegraphics[width=0.9\textwidth]{fig/o2popu.png}
    \caption{Temporal population evolution of $\apiu$, $\bsgm$, $\csup$, $\dsum$, $\esum$, $\fpig$ and $\hsgp$ states of O$_2^+$ cation under different ionization laser intensity at alignment angle $\theta=60$ degrees. From left to right, the laser intensity is $2.2\times 10^{14}$ W/cm$^2$, $1.0\times 10^{14}$ W/cm$^2$ and $0.5\times 10^{14}$ W/cm$^2$, respectively.} 
    \label{fig:o2popu}
\end{figure}
%%%%%%%%%%%%%%%%%%%%%%%%%%%%%%%%%%%%%%

%%%%%%%%%%%%%%%%%%%%%%%%%%%%%%%%%%%%%%%%%
\begin{table}[htb!]
    \centering
    \begin{tabular}{|c|c|}
    \hline
    Channel & $(\theta, E)$-dependence \\
    \hline\hline $\fpig\muzarrow\apiu$ & $E\cos\theta$ \\
    \hline $\fpig\muxyarrow\ \esum\muzarrow\ \hsgp\muxyarrow\apiu$ & $E^3\cos\theta\sin^2\theta$ \\
    $\fpig\muxyarrow\csup\muzarrow\ \hsgp\muxyarrow\apiu$ &  $E^3\cos\theta\sin^2\theta$ \\
    $\fpig\muxyarrow\ \dsum\muzarrow\bsgm\muxyarrow\apiu$ &  $E^3\cos\theta\sin^2\theta$ \\
    \hline
    \end{tabular}
    \caption{The first and third order pathways for population transfer in O$_2^+$ cation, and their dependence on alignment angle and electronic field strength.}
    \label{tab:a2f}
\end{table}
%%%%%%%%%%%%%%%%%%%%%%%%%%%%%%%%%%%%%%%%%

%%%%%%%%%%%%%%%%%%%%%%%%%%%%%%%%%%%%%
%%%%%%%%%%%%%%%%%%%%%%%%%%%%%%%%%%%%%%

\section{Branching ratio of PPRM pathways}

\begin{figure}[htb!]
  \subfloat[]{\includegraphics[width=0.32\textwidth]{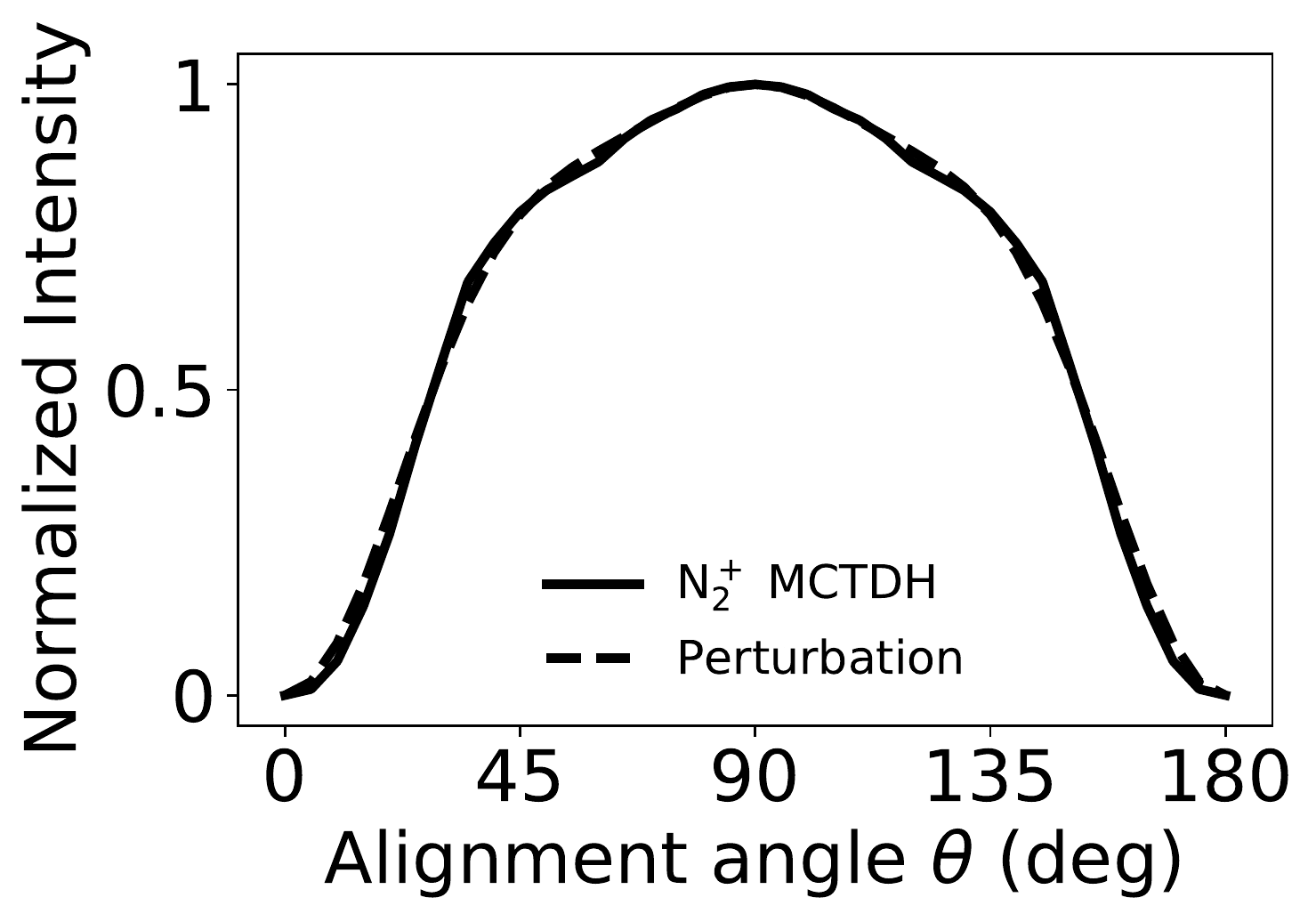}}
 \hfill 	
  \subfloat[]{\includegraphics[width=0.32\textwidth]{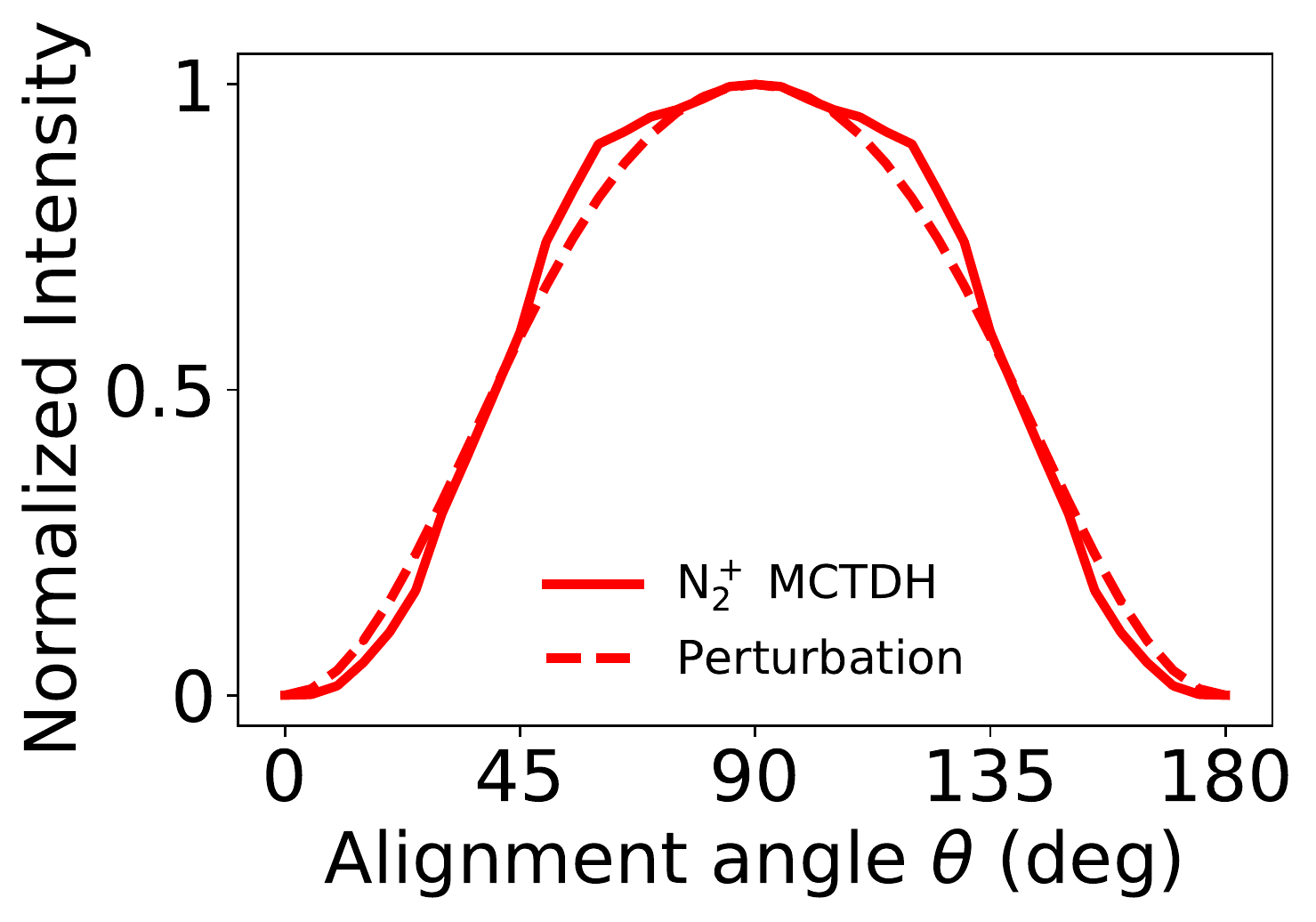}}
 \hfill	
  \subfloat[]{\includegraphics[width=0.32\textwidth]{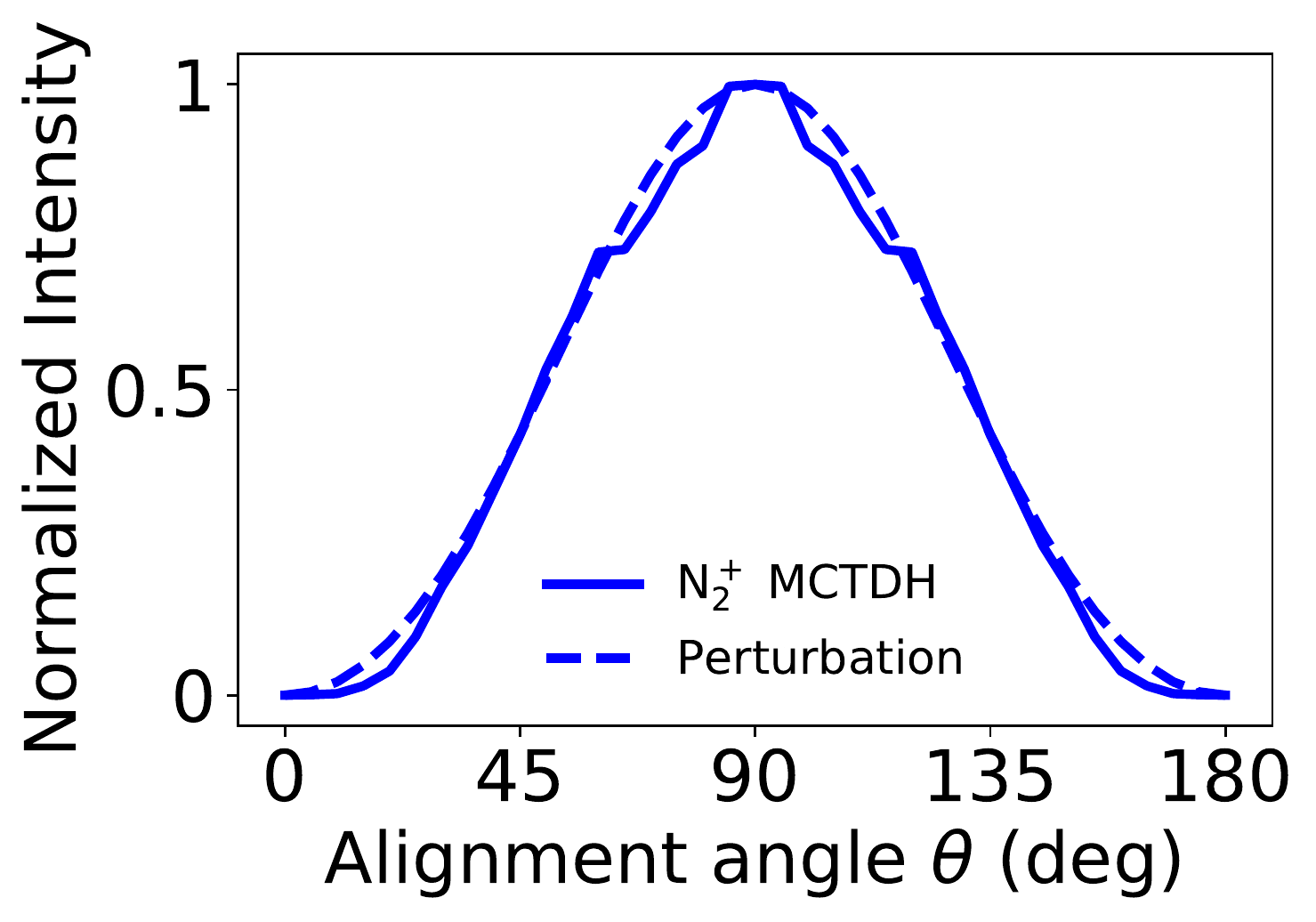}}
  \newline
  \subfloat[]{\includegraphics[width=0.32\textwidth]{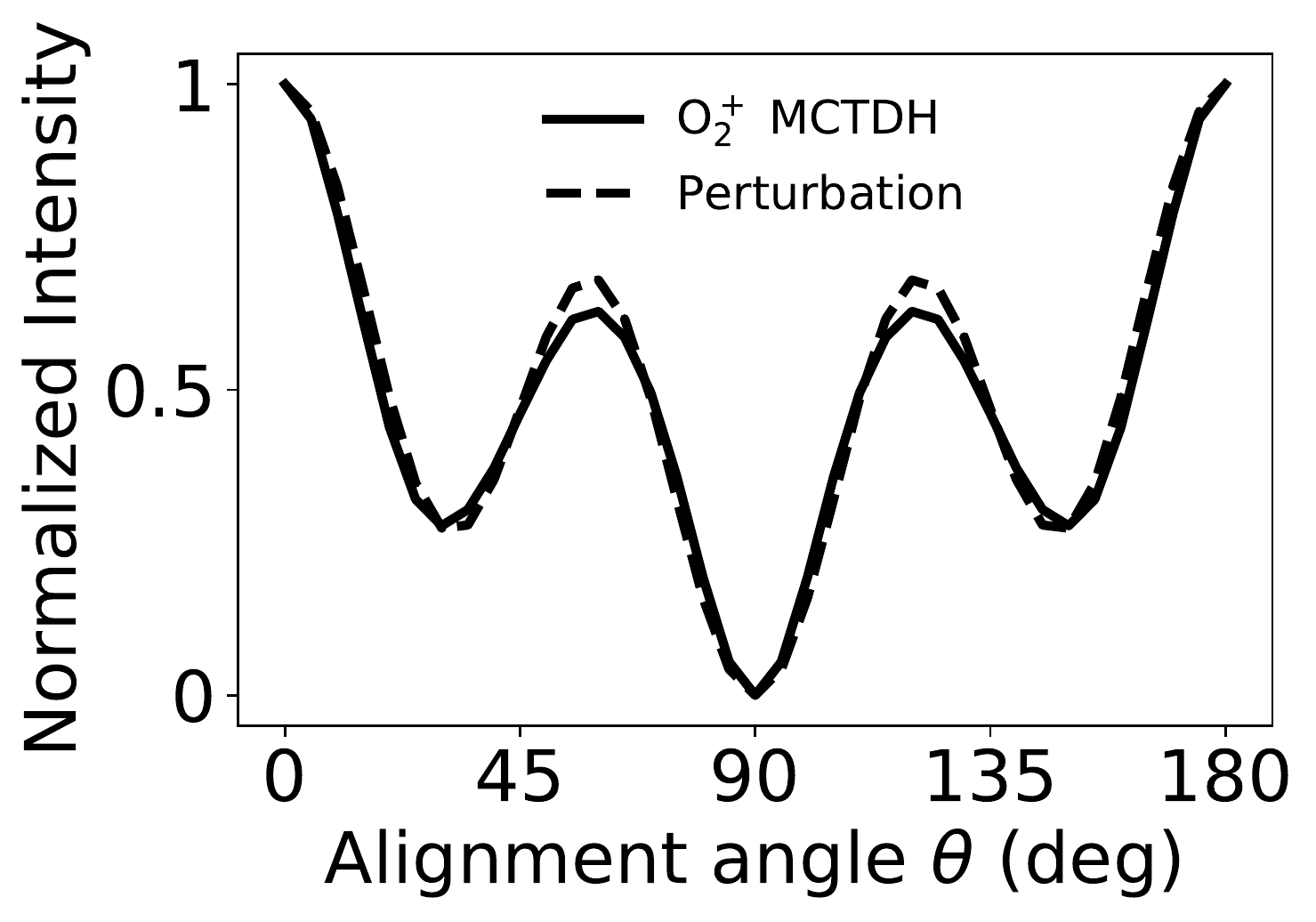}}
 \hfill 	
  \subfloat[]{\includegraphics[width=0.32\textwidth]{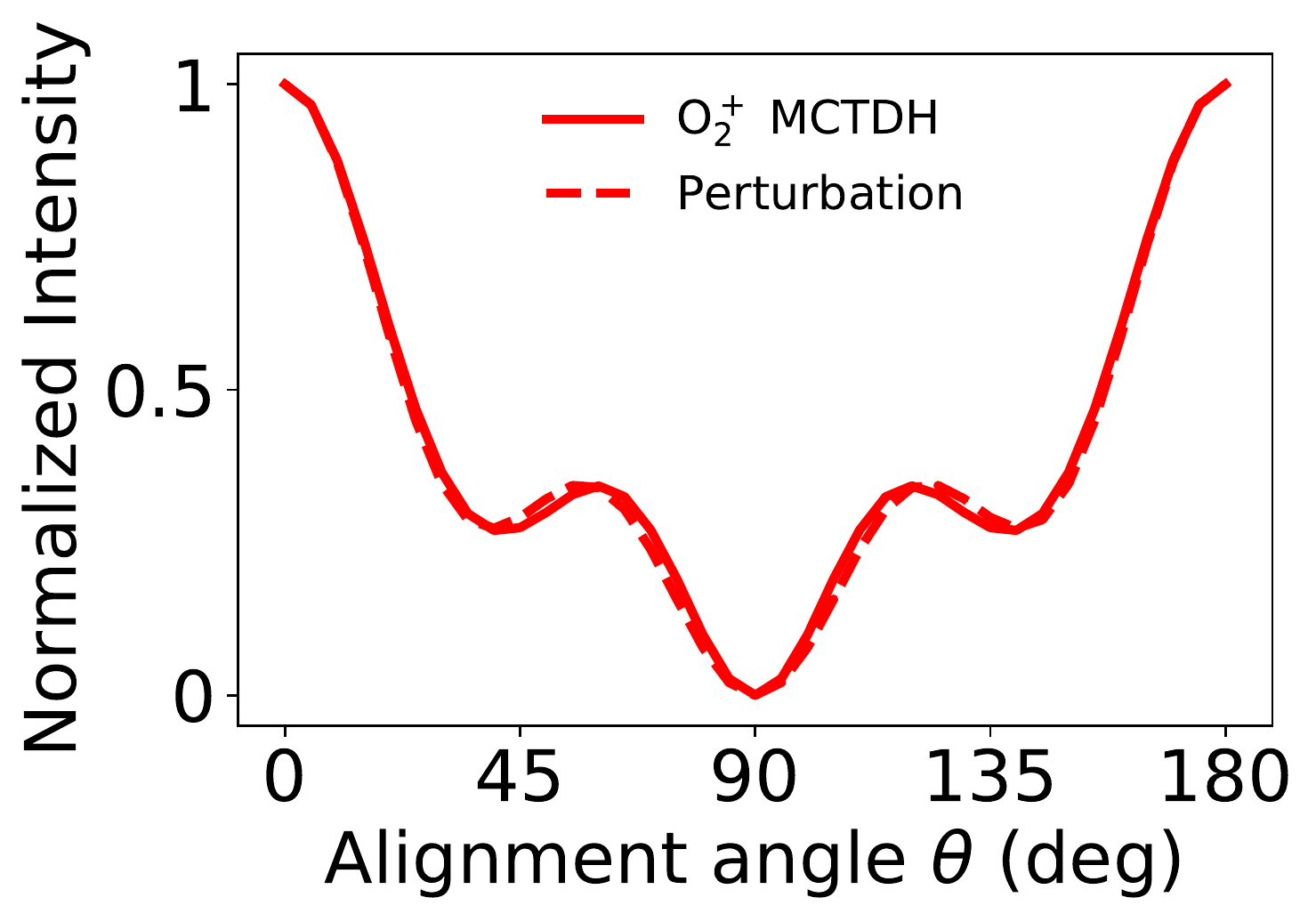}}
 \hfill	
  \subfloat[]{\includegraphics[width=0.32\textwidth]{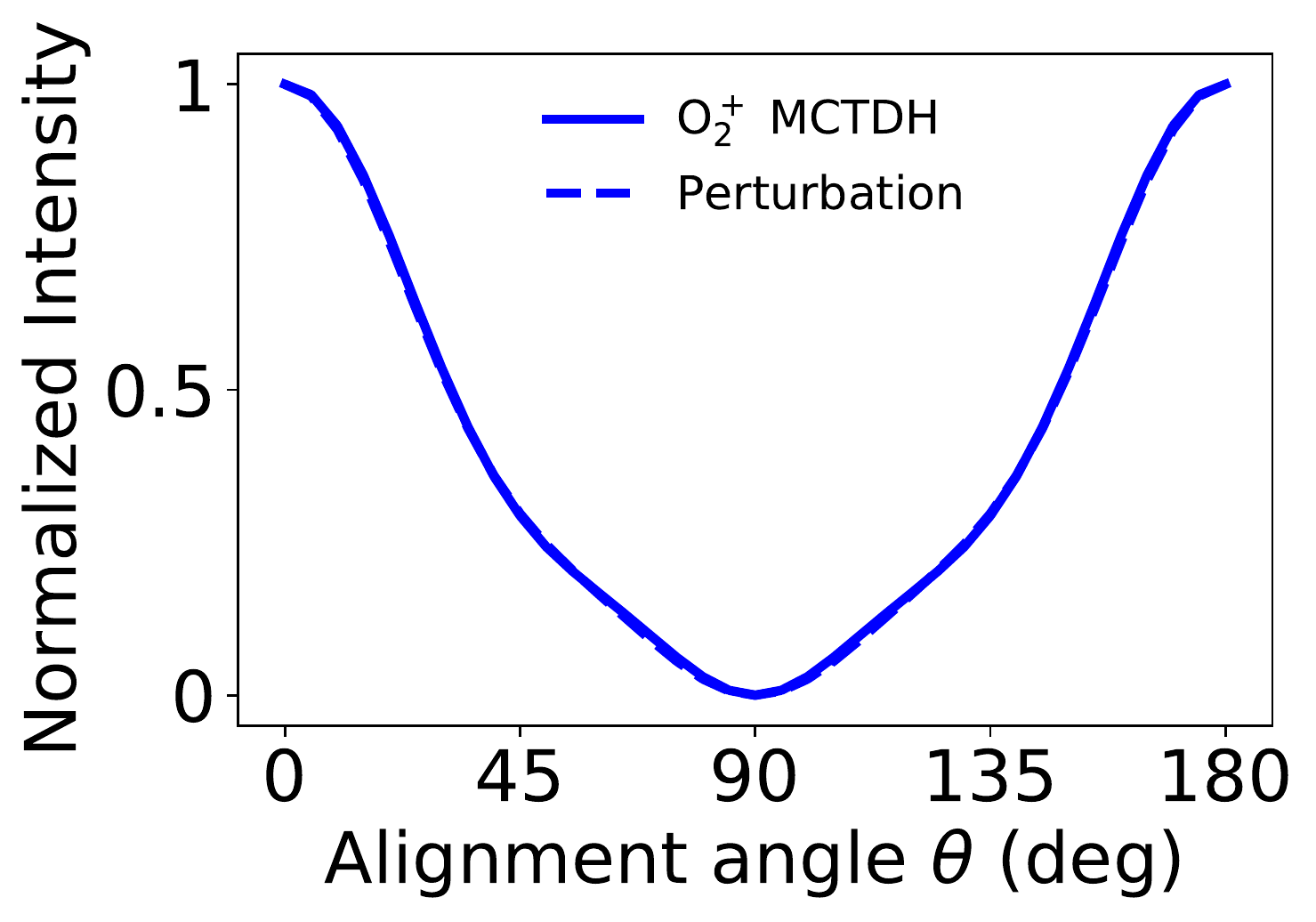}}
    \caption{The upper panel shows the comparison of $\Init$ dissociation yields between MCTDH simulation and the model in Eq.~\ref{eq:fitn2} for laser intensities $2.2\times 10^{14}$ W/cm$^2$, $1.0\times 10^{14}$ W/cm$^2$ and $0.5\times 10^{14}$ W/cm$^2$, respectively. The lower panel shows the comparison of O$_2^+$ dissociation for the same set of intensities. Solid and dashed lines indicate the result of MCTDH simulation and the model, respectively.}
    \label{fig:fit}
\end{figure}

Using symmetry selection rules listed in \tab{tab:X2Piu} and \tab{tab:a2f}, the transitions from cationic ground states to final dissociative states involve only odd order processes. Since the highest laser intensity in our investigation is 2.2$\times10^{14}$ W/cm$^2$, which corresponds to the field strength of 0.075 a.u., processes higher than third order are neglected.
As shown in \fig{fig:fit}, the model consisting of first and third order transitions excellently reproduces the variation of dissociation yields as a function of ionization laser intensity for both N$_2^+$ and O$_2^+$ cation.
%
%%%%%%%%%%%%%%%%%%%%%%%%%%%%%%%%%%%%%%%%
%%%%%%%%%%%%%%%%%%%%%%%%%%%%%%%%%%%%%%
In the model, we fit the the alignment angle-dependent dissociation yields for $\Init$ and O$_2^+$ using the following equation,
%%%%%%%%%%%%%%%%%%%%%%%%%%%%%%%%%%%%%%
\begin{equation}
    D_{\textrm{yields}}(\theta) = A_{0\mathrm{M}}^2 \sin^2\theta + B_{2\mathrm{M}}^2 \sin^2\theta\cos^4\theta+ 2 A_{0\mathrm{M}} B_{2\mathrm{M}} C\sin^2\theta\cos^2\theta\ ,
    \label{eq:fitn2}
\end{equation}
where $A_{0M}$, $B_{2M}$ are transition amplitudes of first and third order transitions with 0 and 2 intermediate states, respectively.
$C=\cos\phi$ represents the effective phase difference $\phi$ between two types of channel. 
For N$_2^+$ cation, though transitions from initial $\Xsgp$ state to the final $^2\Phi_u$ state could also provide minor dissociation yield ($\textless$10\% under highest laser intensity), it has the same alignment angle-dependence as the transitions between $\Xsgp$ and $^2\Pi_u$ states, i.e. the amplitudes are proportional to $E\sin\theta$ for first order transition and proportional to $E^3\cos^2\theta\sin\theta$ for third order transitions.

The branching ratio of first and third order channels for $Init$ and O$_2^+$ and the effective phase difference is shown in \tab{tab:fitn2} and \tab{tab:fito2}, respectively.
%%%%%%%%%%%%%%%%%%%%%%%%%%%%%%%%%%%%
\begin{table}[htb!]
    \centering
    \begin{tabular}{|c|c|c|}
    \hline
    \multirow{2}{*}{Laser Intensity($10^{14}$ W/cm$^2$)} & \multicolumn{2}{c|}{Branching ratio}                                                                       \\ \cline{2-3}
    & \multicolumn{1}{l|}{$\frac{A^2_{0\mathrm{M}}}{A^2_{0\mathrm{M}}+B^2_{2\mathrm{M}}}$} & \multicolumn{1}{l|}{$\frac{B^2_{2\mathrm{M}}}{A^2_{0\mathrm{M}}+B^2_{2\mathrm{M}}}$}    \\ \hline\hline
    2.2                                               & 0.390                                                & 0.610                                                   \\ \hline
    1.0                                               & 0.973                                                & 0.027                                                   \\ \hline
    0.5                                               & 0.978                                                & 0.022                                                   \\ \hline
    \end{tabular}
    \caption{The branching ratio of first and third order pathways of $\Init$ under three ionization laser intensities.}
    \label{tab:fitn2}
\end{table}
%%%%%%%%%%%%%%%%%%%%
%%%%%%%%%%%%%%%%%%%%
\begin{table}[htb!]
    \centering
    \begin{tabular}{|c|c|c|c|}
    \hline
    \multirow{2}{*}{Laser Intensity($10^{14}$ W/cm$^2$)} & \multicolumn{2}{c|}{Branching ratio}                                                                      \\ \cline{2-3}
    & \multicolumn{1}{l|}{$\frac{A^2_{0\mathrm{M}}}{A^2_{0\mathrm{M}}+B^2_{2\mathrm{M}}}$} & \multicolumn{1}{l|}{$\frac{B^2_{2\mathrm{M}}}{A^2_{0\mathrm{M}}+B^2_{2\mathrm{M}}}$}   \\ \hline\hline
    2.2                                               & 0.094                                                & 0.906                                                  \\ \hline
    1.0                                               & 0.160                                                & 0.840                                                  \\ \hline
    0.5                                               & 0.408                                                & 0.592                                                  \\ \hline
    \end{tabular}
    \caption{The branching ratio of first and third order pathways of O$_2^+$ under three ionization laser intensities.}
    \label{tab:fito2}
\end{table}
%%%%%%%%%%%%%%%%%%%%
As expected, the branching ratio of third order channels greatly increases at higher laser intensities, which indicates that PPRM affects the dissociation process by multi-level transitions involving participation of intermediate states.
Because the non-adiabatic couplings between the electronic states are not taken into account, the dissociative states could not receive transferred population after the laser pulse, and the total dissociation ratio could be underestimated. %
However the non-adiabatic coupling does not depend on the alignment angle and its contribution will not affect the analysis of dissociative ionization pathways.

\section{Conclusion}
To conclude, we have performed theoretical investigations for the dissociation of aligned N$_2$ and O$_2$ molecules after ionization by intense short IR laser pulse.
We calculated the alignment angle-resolved kinetic energy release spectra of the dissociated fragments and the variation as a function of ionization laser intensity, from which the branching ratio of first and third order population transfer processes can be determined.
Our method can be conveniently applied in analysis of realistic experiments of strong field ionization of molecules, e.g. for post ionization population redistribution dynamics of N$_2^+$ air laser, and more generally, in analyzing ultrafast population dynamics induced by laser-molecule interaction. 
Theoretical schemes to further disentangle population transfer pathways of multiple initial states and final states is being pursued.

\section{ACKNOWLEDGMENTS}
This work is supported by National Key Research and Development Program of China (Grant No. 2019YFA0308300) and National Natural Science Foundation of China (Grant Nos. 11722432, 12021004, 92050201, and 61475055).

\bibliographystyle{naturemag}
\bibliography{ref}

\end{document}